\documentclass[a4paper,11pt]{article}
\pdfoutput=1 % if your are submitting a pdflatex (i.e. if you have
\usepackage{jheppub} % for details on the use of the package, please
\usepackage[T1]{fontenc} % if needed

\usepackage{cancel}

\newcommand{\SM}{\text{{\large $\mathcal{S}$}}}

\usepackage{enumitem}
\usepackage{mathrsfs}

\newcommand{\BB}{{\scriptscriptstyle\text{B}}}
\newcommand{\FF}{{\scriptscriptstyle\text{F}}}

\newcommand{\de}{\text{d}}

\newcommand{\AdSSSS}{\text{AdS}_3\times\text{S}^3\times\text{S}^3\times\text{S}^1}
\newcommand{\AdSST}{\text{AdS}_3\times\text{S}^3\times\text{T}^4}
\newcommand{\AdSSSSb}{\mathbf{AdS}_{\mathbf{3}}\boldsymbol{\times}\mathbf{S}^{\mathbf{3}}\boldsymbol{\times}\mathbf{S}^{\mathbf{3}}\boldsymbol{\times}\mathbf{S}^{\mathbf{1}}}

\usepackage{tikz}
\usetikzlibrary{math}
\usetikzlibrary{arrows,shapes,positioning}
\usetikzlibrary{decorations.markings}
\tikzstyle arrowstyle=[scale=1]
\tikzstyle directed=[postaction={decorate,decoration={markings,
    mark=at position .65 with {\arrow[arrowstyle]{stealth}}}}]
\tikzstyle reverse directed=[postaction={decorate,decoration={markings,
    mark=at position .65 with {\arrowreversed[arrowstyle]{stealth};}}}]

\newlength{\mywidth}
\newcommand{\bal}{\begin{equation}\begin{aligned}}
\newcommand{\eal}{\end{aligned}\end{equation}}

\newcommand{\ov}{\over}

\newcommand{\B}{{\scriptscriptstyle\text{B}}}
\newcommand{\F}{{\scriptscriptstyle\text{F}}}
\renewcommand{\L}{{\scriptscriptstyle\text{L}}}
\newcommand{\R}{{\scriptscriptstyle\text{R}}}

\def\cX{{\cal X}}
\def\tx{{\tilde x}}
\def\tg{{\tilde\gamma}}
\def\bes{{\text{\tiny BES}}}

\def\hl{{\text{\tiny HL}}}
\def\tp{{\widetilde p}}
\newcommand{\E}{\mathcal E}
\def\tE{\widetilde\E}

\title{\boldmath Relativistic limit on the $AdS_3 \times S^3 \times S^3 \times S^1$ string worldsheet}

\author[a]{Emanuele Maria Cattaneo,}
\author[b]{Davide Polvara,}
\author[c,d,1]{Alessandro Sfondrini\note{On leave from the University of Padova, Italy.}}

\affiliation[a]{Dipartimento di Fisica e Astronomia, Universit\`a degli Studi di Padova, via Marzolo 8,
35131 Padova, Italy.}
\affiliation[b]{II. Institut f\"ur Theoretische Physik,  Universit\"at Hamburg, Luruper Chaussee 149, 22761 Hamburg, Germany.}
\affiliation[c]{School of Mathematics, University of Birmingham, Edgbaston B15 2TT, UK.}
\affiliation[d]{Istituto Nazionale di Fisica Nucleare, Sezione di Padova, via Marzolo 8, 35131 Padova,
Italy.}

\emailAdd{emanuelemaria.cattaneo@studenti.unipd.it}
\emailAdd{davide.polvara@gmail.com}
\emailAdd{a.sfondrini@bham.ac.uk}

\abstract{
We study the relativistic limit of the worldsheet S-matrix of $AdS_3 \times S^3 \times S^3 \times S^1$ strings in the presence of mixed Ramond–Ramond (RR) and Neveu-Schwarz–Neveu-Schwarz (NS-NS) flux and complete the S-matrix bootstrap including the dressing factors for fundamental particles and bound states. Our results agree with the relativistic limit of the recent proposal arXiv:2512.07721. 
}

\begin{document} \begin{flushright}\small{ZMP-HH/26-1}\end{flushright}
\maketitle
\flushbottom

%%%%%%%%%%%%%%%%%%%%%%%%%%%%
\section{Introduction}
Superstring theory of $\AdSSSS$ is an important playground for the holographic correspondence since its inception~\cite{Maldacena:1997re,Elitzur:1998mm,Boonstra:1998yu}. Like it is the case for the arguably simpler $\AdSST$ background, the most powerful approach to qualitatively understand the string dynamics in the presence of Ramond--Ramond (RR) fluxes is worldsheet integrability.%
\footnote{%
The integrability approach was first developed for $\text{AdS}_5\times\text{S}^5$ superstrings and the dual $\mathcal{N}=4$ supersymmetric Yang--Mills (SYM) theory, see~\cite{Arutyunov:2009ga, Beisert:2010jr} for reviews and references.}
Classical integrability was established for $\AdSSSS$ strings supported by RR flux only in~\cite{Babichenko:2009dk}, and for backgrounds supported by a mixture of RR and Neveu--Schwarz--Neveu--Schwarz (NSNS) flux in~\cite{Cagnazzo:2012se}.
An investigation of the symmetries of the model was then carried out~\cite{OhlssonSax:2011ms, Borsato:2012ud,Borsato:2012ss,Borsato:2015mma}, in analogy with what had been done for $\mathcal{N}=4$ SYM~\cite{Beisert:2005tm} and $\text{AdS}_5\times\text{S}^5$ superstrings~\cite{Arutyunov:2006ak,Arutyunov:2006yd}. This, along with the perturbative and semiclassical investigation of the worldsheet model~\cite{Rughoonauth:2012qd,Sundin:2012gc,Sundin:2013ypa,Hoare:2013lja,Bianchi:2014rfa} allowed to determine \textit{almost entirely} the worldsheet S~matrix of the model, which was proposed in 2015 in~\cite{Borsato:2015mma}.
While this was sufficient to elucidate some properties of the model, such as the protected spectrum~\cite{Baggio:2017kza}, lacking the complete knowledge of the S~matrix meant that a quantitative study of the non-protected spectrum stalled. 
More precisely, the S~matrix was determined \textit{up to its dressing factors} --- multiplicative pre-factor of the S~matrix which cannot be fixed by linearly realised symmetry, but should be determined based on unitarity, crossing symmetry~\cite{Janik:2006dc} and ``good analytic behavior''. This is the case for all integrable quantum field theories. For relativistic models, there is usually a single dressing factor out of which all others can be constructed, and it is a meromorphic function on the $\theta$-rapidity plane.%
\footnote{In a two-dimensional relativistic QFT, one introduce rapidities $\theta_j$ so that the momenta are $p_j=m_j \sinh\theta_j$ and the energies are $H_j=m_j \cosh\theta_j$. Then, the two-particle S~matrix depends only on the difference $\theta\equiv\theta_1-\theta_2$.}
Because string worldsheet models are non-relativistic (as a consequence of the lightcone gauge-fixing) the dressing factors are substantially more involved. For instance, for $\text{AdS}_5\times\text{S}^5$ superstrings there is also a single dressing factor, the celebrated Beisert--Eden--Staudacher (BES) function~\cite{Beisert:2006ib}, which has quite nontrivial analytic properties~\cite{Dorey:2007xn,Arutyunov:2009kf}.
For $\text{AdS}_3$ integrable superstrings backgrounds the situation is more complicated, and only recently the dressing factors had been proposed --- first for RR-only backgrounds~\cite{Frolov:2021fmj} and then for mixed RR-NSNS flux ones~\cite{Frolov:2024pkz,Frolov:2025uwz,Frolov:2025tda}.%
\footnote{One may wonder what happens for pure-NSNS backgrounds. In that case, the worldsheet S~matrix is entirely given by a ``CDD''~\cite{Castillejo:1955ed} factor~\cite{Baggio:2018gct}, and the integrability construction matches the worldsheet-CFT prediction~\cite{Maldacena:2000hw}. This was worked out explicitly for $\AdSST$~\cite{Dei:2018mfl} and $\AdSSSS$ superstrings~\cite{Dei:2018jyj}.
}

Only in the last few months, efforts turned again to the study of the~$\AdSSSS$ background~\cite{Cavaglia:2025icd,Chernikov:2025jko,Frolov:2025syj}. Before discussing these advances, let us review the features of this background and of its integrability description, which is more complex than that of $\text{AdS}_5\times\text{S}^5$ and $\AdSST$. In the case of mixed RR and NSNS flux, where we normalise the radius of $\text{AdS}_3$ to one, the NSNS flux~$H_3$ and RR three-form flux $F_3$  can be taken to be
\begin{equation}
\de B=H_3= 2 q\, \Omega\,,\qquad
F_3= 2 \sqrt{1-q^2}\,\Omega\,,\qquad
 \,,\qquad0\leq q \leq 1\,,
\end{equation}
where $B$ is the Kalb--Ramond field, the other fluxes vanish, and we introduced the volume form 
\begin{equation}
    \Omega=\text{Vol}(\text{AdS}_3) + R^2_{(1)} \text{Vol}(\text{S}_{(1)}^3)+ R^2_{(2)} \text{Vol}(\text{S}_{(2)}^3)\,.
\end{equation}
Here we stripped out the radii of the two three-spheres $R_{(1)}$ and $R_{(2)}$, so that the volume forms are for unit-radius spheres. The supergravity equations require that
\begin{equation}
    1=\frac{1}{R^2_{(1)}}+\frac{1}{R^2_{(2)}}\qquad\Rightarrow
    \qquad\alpha\equiv\frac{1}{R^2_{(1)}}\,,\quad 1-\alpha=\frac{1}{R^2_{(2)}}\,,\qquad0\leq\alpha\leq1\,.
\end{equation}
The quantisation of the Wess--Zumino term for the two three-spheres in the string action results in the quantisation conditions
\begin{equation}
T \int\limits_{\text{S}^3_{(1)}} H_3 = 2 \pi k_1 \,, \qquad T \int\limits_{\text{S}^3_{(2)}} H_3 = 2 \pi k_2\,, \qquad k_1\,, k_2 \in \mathbb{N} \,,
\end{equation}
where $T>0$ is the string tension (in units where $R_{\text{AdS}_3}=1$). This results in the conditions
\begin{equation}
    \frac{qT}{\alpha}=\frac{k_1}{2\pi}\,,\qquad
    \frac{qT}{1-\alpha}=\frac{k_2}{2\pi}\,.
\end{equation}
It is convenient to re-package the parameters determining the background in terms of the NSNS fluxes through the spheres $k_1,k_2\in\mathbb{N}$ and the amount of RR flux $h=\sqrt{1-q^2}T\geq0$, which is continuous in perturbative string theory and corresponds to a marginal coupling in the dual CFT$_2$. In terms of those
\begin{equation}
\label{eq:alphakmap}
    \alpha=\frac{k_2}{k_1+k_2}\,,\qquad
    1-\alpha=\frac{k_1}{k_1+k_2}\,,\qquad
    T=\sqrt{h^2+\frac{k^2}{4\pi^2}}\,,\qquad
    k=\frac{k_1k_2}{k_1+k_2}\,.
\end{equation}
In the case where $h=0$ and the model admits a worldsheet-CFT description~\cite{Maldacena:2000hw}, $k$ is related to the level of an $\mathfrak{sl}(2)_k$ Ka\v{c}--Moody algebra --- see e.g.~\cite{Dei:2018yth} for a detailed discussion of the parameters and conventions.  Note that, in contrast with the case of $\AdSST$, here $k$ need not be integer. We finally note that we can formally recover the case of $\AdSST$ (or more precisely, $\text{AdS}_3\times \text{S}^3\times \mathbb{R}^3\times \text{S}^1$) by blowing up one of the spheres, i.e.\ $k_1\to\infty$ or $k_2\to\infty$.

Let us now come back to the very recent advances in the study of the $\AdSSSS$ backgrounds~\cite{Cavaglia:2025icd,Chernikov:2025jko,Frolov:2025syj}.
In two independent works~\cite{Cavaglia:2025icd,Chernikov:2025jko} the Quantum Spectral Curve (QSC) for the $\AdSSSS$ background was proposed. Both works focus on pure-RR backgrounds and on the case $\alpha=1/2$ (though~\cite{Cavaglia:2025icd} also comments on pure-RR backgrounds for general~$\alpha$). The QSC in principle follows from the S-matrix bootstrap through the derivation of the ``mirror''~\cite{Arutyunov:2007tc} thermodynamic Bethe ansatz (TBA), Y-system, and T-Q relations. In the case at hand, however, the QSC was directly conjectured based on symmetries and analyticity. Regardless, the QSC contains in principle the information about the S~matrix, inducing the dressing factors, though it can be involved to extract it. Both  works~\cite{Cavaglia:2025icd,Chernikov:2025jko} found that the QSC construction was incompatible with requiring both crossing invariance and braiding unitarity --- quite surprisingly, and in contrast to all previously known worldsheet integrability setups. 
Shortly afterwards, it was pointed out in~\cite{Frolov:2025syj} that it was possible to construct crossing symmetric dressing factors with self-consistent pole structure and which obey braiding unitarity. Moreover, the solution can be found for any $k_1,k_2$ and $h$.

Given these developments, the aim of this work is to test the proposal of~\cite{Frolov:2025syj}, and discuss whether other proposals with reasonable analytic properties are possible. The basic checks of~\cite{Frolov:2025syj} already include the near-BMN~\cite{Berenstein:2002jq} expansion of the dressing factors, which matches the results in~\cite{Bianchi:2014rfa}.
Here we consider a different limit, which is possible for mixed-flux $\text{AdS}_3$ worldsheet models, whereby \textit{the worldsheet model becomes relativistic}. In this limit it is significantly easier to construct the S~matrix, complete with its dressing factors, by the usual S-matrix bootstrap, and to check that the bound-state fusion closes in a self-consistent manner.%
\footnote{For an introduction to fusion in relativistic integrable QFTs see e.g.~\cite{Dorey:1996gd}.}
This relativistic S~matrix becomes a benchmark for (the limit of) the dressing factors of the original model, in this case of~\cite{Frolov:2025syj}. This relativistic limit was already studied for mixed-flux $\AdSST$ case in~\cite{Frolov:2023lwd}. The resulting relativistic S~matrix turned out to be closely related to the one worked out by Fendley and Intriligator in the 1990s~\cite{Fendley:1992dm}.%
\footnote{A similar relativistic limit was also considered in~\cite{Fontanella:2019ury}, though with a different interpretation and particle content with respect to~\cite{Frolov:2023lwd}.}
It was then checked~\cite{Frolov:2025uwz} that the full (non-relativistic) dressing factor of mixed-flux $\AdSST$ strings correctly reproduces the relativistic limit.
In a nutshell, here we want to do the same for mixed-flux $\AdSSSS$ strings.

This paper is structured as follows. We begin by briefly reviewing the particle content and symmetries of $\AdSSSS$ in the lightcone gauge in section~\ref{sec:reviewSmat}.
In section~\ref{sec:limit} we work out the relativistic limit and construct the relativistic S~matrix, and discuss the comparison with the recent proposal~\cite{Frolov:2025uwz}. Perhaps unsurprisingly, the limit closely resembles that of $\AdSST$~\cite{Frolov:2023lwd} and the S~matrix closely relates to that of Fendley--Intriligator~\cite{Fendley:1992dm}. For this reason, we summarise that literature in appendix~\ref{app:reviewlimit}.
We conclude in section~\ref{sec:conclusions}, and in appendix~\ref{app:example} we work out a specific example ($k_1=3$, $k_2=6$) for the readers' convenience.

%%%%%%%%%%%%%%%%%%%%%%%%%%%%
\section{String on the mixed-flux $\AdSSSSb$ background}
\label{sec:reviewSmat}

We consider the background $\AdSSSS$, following the conventions of~\cite{Borsato:2015mma}. 
To distinguish the two three-spheres, we use labels 1 and~2.

\subsection{Symmetries in ligthcone gauge}

The supersymmetry algebra of $\AdSSSS$ is given by two copies of $\mathfrak{d}(2,1; \alpha)$ exceptional Lie superalgebra, which we denote by left and right, $\mathfrak{d}(2,1; \alpha)_{\L} \oplus \mathfrak{d}(2,1; \alpha)_{\R}$. The parameter $\alpha$ is precisely the one from eq.~\eqref{eq:alphakmap}.
After light-cone gauge fixing~\cite{Borsato:2015mma}, the linearly realised symmetries are
\begin{equation}
\label{eq:lcalgebra}
(\mathfrak{su}(1|1)_\L \oplus \mathfrak{su}(1|1)_\R)_{c.e.} \subset \mathfrak{d}(2,1; \alpha)_{\L} \oplus \mathfrak{d}(2,1; \alpha)_{\R} \,,
\end{equation}
where the subscript ``c.e.'' indicates two central extensions $\mathbf{C}$ and $\mathbf{\bar{C}}$ (which are conjugate to each other on unitary representations). More specifically, the algebra~\eqref{eq:lcalgebra} is defined by the anticommutation relations
\begin{equation}
\label{eq:lcalgebracomm}
\begin{aligned}
    &\{ \mathbf{Q},  \mathbf{S} \}= \frac{1}{2} (\mathbf{H}+\mathbf{M}),\qquad &&\{ \mathbf{Q},  \mathbf{\tilde{Q}} \}= \mathbf{C} \; ,\\
    &\{ \mathbf{\tilde{Q}},  \mathbf{\tilde{S}} \}= \frac{1}{2} (\mathbf{H}-\mathbf{M}),\qquad &&\{ \mathbf{S},  \mathbf{\tilde{S}} \}= \mathbf{\bar{C}} .
\end{aligned}
\end{equation}
Here $\mathbf{Q},\mathbf{S}$ generate $\mathfrak{su}(1|1)_\L$ and $\tilde{\mathbf{Q}},\tilde{\mathbf{S}}$ generate $\mathfrak{su}(1|1)_\R$. Notice that only \textit{one quarter} of the original sixteen odd generators of $\mathfrak{d}(2,1; \alpha)_{\L} \oplus \mathfrak{d}(2,1; \alpha)_{\R}$ have survived the lightcone gauge fixing. This is in contrast with the case of $\AdSST$ where \textit{one half} of the odd generators survive. 
The central extensions $\mathbf{C} ,\mathbf{\bar{C}} $ are a unique feature of the lightcone gauge-fixed model~\cite{Arutyunov:2006ak,Borsato:2015mma}, similar to Beisert's central extension~\cite{Beisert:2005tm}. The charges $\mathbf{H},\mathbf{M}$ are linear combinations of the original Cartan elements of $\mathfrak{d}(2,1; \alpha)_{\L} \oplus \mathfrak{d}(2,1; \alpha)_{\R}$. To see this, let us call $\mathbf{L}_0,\mathbf{\tilde{L}}_0$ the Cartan elements of $\mathfrak{so}(2,2)$, and $\mathbf{J}^3_{(j)},\mathbf{\tilde{J}}^3_{(j)}$ the Cartan elements of either $\mathfrak{so}(4)$, distinguished by $j=1,2$. 
For each copy of $\mathfrak{d}(2,1;\alpha)$ we consider the orthogonal combinations
\begin{equation}
\begin{aligned}
    \mathcal{E}&=\mathbf{L}_0- \alpha\,\mathbf{J}^3_{(1)}-(1-\alpha)\, \mathbf{J}^3_{(2)}\,,\\
    \mathcal{P}&=\mathbf{L}_0+ \alpha\,\mathbf{J}^3_{(1)}+(1-\alpha)\, \mathbf{J}^3_{(2)}\,,\\
    \mathcal{D}&=\phantom{\mathbf{L}_0}+\phantom{\alpha}\,\mathbf{J}^3_{(1)}-\phantom{(1-\alpha)}\, \mathbf{J}^3_{(2)}\,,
\end{aligned}
\end{equation}
and similarly~$\tilde{\mathcal{E}},\tilde{\mathcal{P}},\tilde{\mathcal{D}}$. By the BPS bound of $\mathfrak{d}(2,1; \alpha)_{\L}$ and $\mathfrak{d}(2,1; \alpha)_{\R}$ we have that on the highest-weight state of unitary representations
\begin{equation}
    \mathcal{E}\geq0\,,\qquad\tilde{\mathcal{E}}\geq0\,,
\end{equation}
and the equality holds only for short (atypical) representations of $\mathfrak{d}(2,1; \alpha)_{\L}$ or $\mathfrak{d}(2,1; \alpha)_{\R}$ (or both), respectively.
For the purpose of computing the S-matrix, one takes a decompactification limit whereby
\begin{equation}
    \mathcal{P}+\tilde{\mathcal{P}}\to\infty\,,
\end{equation}
so that one generator decouples. As for the remaining ones, we have 
\begin{equation}
\label{eq:u1charges}
\mathbf{H}=\mathcal{E}+\tilde{\mathcal{E}}\,,\qquad
\mathbf{M}=\mathcal{E}-\tilde{\mathcal{E}}\,,\qquad
\mathcal{D}\pm\tilde{\mathcal{D}}\,,\qquad
\mathcal{P}-\tilde{\mathcal{P}}\,.
\end{equation}
As we highlighted, the first two combinations are those appearing in~\eqref{eq:lcalgebracomm}. In particular, $\mathbf{H}\geq0$ is precisely the lightcone Hamiltonian. The three remaining linear combinations act as automorphisms on~\eqref{eq:lcalgebracomm} and can be used to distinguish different particles with the same eigenvalues of $\mathbf{H}$, $\mathbf{M}$, and $\mathbf{C}, \mathbf{\bar{C}}$.

\paragraph{Representations and particles.}
Excitations transform in irreducible short representations of the lightcone symmetry algebra on which $\mathbf{H}$, $\mathbf{M}$, $\mathbf{C}$ and $\mathbf{\bar{C}}$ have eigenvalues~\cite{Hoare:2013lja,Lloyd:2014bsa,Borsato:2015mma}
\begin{equation}
\label{eq:reprEigen}
\begin{aligned}
&H_{m}(p)=\sqrt{\Bigl(m+\frac{k}{2 \pi}p \Bigr)^2 +4 h^2 \sin^2 \frac{p}{2}} \, ,\qquad
        &&C_m(p)= +\frac{i h}{2} \bigl( e^{+ip}-1 \bigr) \, ,\\
& M_{m}(p)=m+\frac{k}{2 \pi}p\, ,\qquad
        &&\bar{C}_m(p)= -\frac{i h}{2} \bigl( e^{-ip}-1 \bigr) \,.
\end{aligned}
\end{equation}
Here $p$ is the worldsheet momentum of the particle, and $m$ distinguishes the type of particle. In particular, we are interested in the values $m= \pm \alpha$ and $m= \pm (1- \alpha)$, corresponding to representations containing a boson from sphere~1 ($|m|=\alpha$) or from sphere~2 ($|m|=1-\alpha$). As it turns out, short representations of~\eqref{eq:lcalgebra} are two-dimensional, comprising a boson and a fermion, and the values $|m|=\alpha$ and $|m|=1-\alpha$ are the smallest possible values of $|m|>0$.
This is relevant for our next consideration because, upon fusion, we expect $m$ to be additive due to the shortening condition~\cite{Hoare:2013lja,Lloyd:2014bsa,Borsato:2015mma}.

To fix the notation and for future reference let us introduce representations $\rho_\star^\BB(m,p)$, having a \textit{bosonic} highest weight state (HWS), mass $m$ and momentum $p$, as well as representations $\rho_\star^\FF(m,p)$, having a \textit{fermionic} highest weight state (HWS), mass $m$ and momentum $p$. The label $\star$ can either be L (left)  or R (right). It corresponds to the parameterisation used for the representation in terms of Zhukovsky variables and plays an important role in the crossing symmetry of this model before the limit~\cite{Borsato:2015mma}.

%More specifically, we adopt the following notation.
%We label by $\phi$ the highest weight state (HWS) and by $\varphi$ the lowest weight state (LWS). They can be either bosons or fermions (naturally if $\phi$ is a boson then $\varphi$ is a fermion and vice-versa) and are identified by the action of the supercharges. Below, we write the action of the supercharges on these representations. 

%%%%%
\paragraph{Representations with bosonic highest-weight state, $\rho^\BB_\star (m,p)$.}
Denoting with $\phi^\B$ the h.w.s.\ and with $\varphi^\F$ the l.w.s., we have
\begin{equation}
\label{eq:repr-bosonic}
\begin{aligned}
     &\mathbf{Q} \, |\phi^\B (p)\rangle = a_m (p)  \,|\varphi^\F(p)\rangle\,, \qquad &&\mathbf{S} \, |\varphi^\F(p)\rangle = \bar{a}_m (p) \, |\phi^\B(p)\rangle ,\\
     &\mathbf{\tilde{Q}} \, |\varphi^\F(p)\rangle = b_m( p) \,|\phi^\B(p)\rangle\,, \qquad &&\mathbf{\tilde{S}} \, |\phi^\B(p)\rangle = \bar{b}_m (p)  \,|\varphi^\F(p)\rangle.
\end{aligned}
\end{equation}

%%%%%
\paragraph{Representations with bosonic highest-weight state, $\rho^\F_\star (m,p)$.}
Denoting with $\phi^\F$ the h.w.s.\ and with $\varphi^\B$ the l.w.s., we have a representation which is almost identical to~\eqref{eq:repr-bosonic} up to changing the statistics, i.e.\ swappign~B$\leftrightarrow$F:
\begin{equation}
\label{eq:repr-fermionic}
\begin{aligned}
     &\mathbf{Q} \, |\phi^\F (p)\rangle = a_m (p)  \,|\varphi^\B(p)\rangle\,, \qquad &&\mathbf{S} \, |\varphi^\B(p)\rangle = \bar{a}_m (p) \, |\phi^\F(p)\rangle ,\\
     &\mathbf{\tilde{Q}} \, |\varphi^\B(p)\rangle = b_m( p) \,|\phi^\F(p)\rangle\,, \qquad &&\mathbf{\tilde{S}} \, |\phi^\F(p)\rangle = \bar{b}_m (p)  \,|\varphi^\B(p)\rangle.
\end{aligned}
\end{equation}

The representation coefficients would strictly speaking depend (through the Zhukovsky variables) on whether the representation is right or left, but keep this dependence implicit. They are fixed (up to an inconsequential phase factor) by using the relations~\eqref{eq:lcalgebracomm} and imposing agreement with~\eqref{eq:reprEigen}. Algebraically, this can be done for any~$m\in\mathbb{R}$, but agreement with the original string model restricts the possible values of $m$. Comparing with the near-BMN expansion of the string action~\cite{Borsato:2015mma} one finds the eight options
\begin{equation}
m= 0\,,\ 0\,, \ \pm \alpha\,, \ \pm (1-\alpha)\,, \ \pm 1 \,,
\end{equation}
with $\alpha$ as in~\eqref{eq:alphakmap}. In what follows, it feels natural (though strictly speaking slightly improper) to call $m$ ``mass''. 
They can be fit into two-dimensional representations as follows.

%%%%
\paragraph{Sphere-1 bosons and their superpartners.}

The lightcone gauge-fixing coordinate is a linear combination of the coordinates on the equator of each sphere and of time in AdS$_3$~\cite{Borsato:2015mma}. Calling the bosons from the first sphere that are not involved in the lighcone gauge-fixing $Y,\,\bar{Y}$, we have
\bal
(Y,\, \psi) \in \rho_{\L}^{\B}(+\alpha,p)\,, \qquad (\bar{\psi},\, \bar{Y}) \in \rho_{\R}^{\F}(-\alpha,p) \,.
\eal

%%%%
\paragraph{Sphere-2 bosons and their superpartners.}

Similarly, calling the bosons from the second sphere that are not involved in the gauge-fixing $X,\,\bar{X}$, we have
\bal
(X, \chi) \in \rho_{\L}^{\B}(+1-\alpha,p)\,, \qquad (\bar{\chi}, \bar{X}) \in \rho_{\R}^{\F}(-1+\alpha,p) \,.
\eal
%%%%
\paragraph{AdS-bosons and their superpatrners.}

Denoting the bosons from AdS$_3$ distinct from the global-time coordinate as $Z,\,\bar{Z}$, these excitations have mass $m=\pm1$. For this reason, it is natural to wonder whether in the full quantum worldsheet theory (at a finite value of the string tension), these excitations should be regarded as bound-state or composite modes. It appears that the latter is more likely~\cite{Sundin:2012gc}, similar to what happens for $\text{AdS}_4\times \mathbb{C}\text{P}^3$ strings~\cite{Zarembo:2009au}. For this reason, these representations will not feature much in our discussion. They are labeled as
\bal
\label{eq:adsboson}
(\vartheta, Z) \in \rho_{\L}^{\F}(+1,p)\,, \qquad (\bar{Z}, \bar{\vartheta}) \in \rho_{\R}^{\B}(-1,p) \,.
\eal

%%%%
\paragraph{``Massless'' particles.}

We finally have two bosons with $m=0$ and their superpartners. One bosons emerges from the $S^1$ factor in $\AdSSSS$, while the other is a linear combination of the coordinates on the equator of each sphere which is orthogonal to the gauge-fixing coordinate, see~\cite{Dei:2018yth}. We find
\bal
\label{eq:masslessrep}
(T, \zeta) \in \rho_{\L}^{\B}(0,p)\,, \qquad (\bar{\zeta}, \bar{T}) \in \rho_{\R}^{\F}(0,p) \,.
\eal

The crossing transformations swap $L\to R$ and representations with $m$ to ones with $-m$. For the special case~$m=0$, it maps the two representations~\eqref{eq:masslessrep} into each other~\cite{Borsato:2015mma}.

\paragraph{Isomorphisms between left and right representations.}
Looking at the central charges in~\eqref{eq:reprEigen} we see that they are left invariant by the simultaneous shift
\begin{equation}
    m\to m+N\,k\,,\qquad p\to p-2\pi\,N\,,\qquad N\in\mathbb{Z}\,.
\end{equation}
This fact was observed first for~$\AdSST$, as discussed at length in~\cite{Frolov:2023lwd}, but it also holds for $\AdSSSS$. It is worth noting that here $m\to m+N k$ means, for instance
\begin{equation}
    \alpha=\frac{k_2}{k_1+k_2}\to \frac{k_2}{k_1+k_2}+N\,k=
    \frac{k_2}{k_1+k_2}+\frac{N\,k_1k_2}{k_1+k_2}=\alpha\,(1+N\,k_1)\,,
\end{equation}
and similarly for $(1-\alpha)$.
Keeping into account the coproduct on multi-particle representations as discussed in~\cite{Frolov:2023lwd}, this corresponds to the following isomorphisms, valid for $0<p<2\pi$ and $0<m<k$,
\begin{equation}
    \rho_\L^\B(m+|N|k,p-2\pi|N|)\cong
    \rho_\L^\B(m,p)\,,\qquad
    \rho_\R^\F(-m-|N|k,p+2\pi|N|)\cong
    \rho_\R^\F(-m,p)\,,
\end{equation}
and
\begin{equation}
\label{eq:lrisomorphism}
    \rho_\L^\B(k-m,p+2\pi)\cong \rho_\R^\F(-m,p)\,,\qquad
    \rho_\R^\F(m-k,p-2\pi)\cong \rho_\L^\B(m,p)\,.
\end{equation}
While both of these equations yield identities for the matrix part of the S~matrix, it turns out that the map~\eqref{eq:lrisomorphism} yields a monodromy in the dressing factors, at least in the case of $\AdSST$.\footnote{See in particular equation (5.56) in~\cite{Frolov:2025uwz}.}
Nonetheless, we will see below that a similar identity holds  in the relativistic limit without any monodromy factor, as it was already observed for $\AdSST$~\cite{Frolov:2025uwz}.

%%%%%%%%%%%%%%%%%%%%%%%%%%
\subsection{Expected fusion in the string and mirror models}

The model's fusion structure was conjectured in~\cite{Frolov:2025syj} based on analyticity considerations and the requirement that the S~matrix should project on short bound-state representations. Let us briefly review that conjecture.

%%%%
\paragraph{Fusion in the string model.}
As argued in~\cite{Frolov:2025syj}, the semiclassical arguments valid for $\text{AdS}_5\times\text{S}^5$ and $\AdSST$ suprestrings suggest that bosonic particles living on the same sphere should fuse together, in the string kinematics. For $\AdSSSS$, we expect that in the string model particles of the same mass $|m|$ (which can be either $\alpha$ or $1-\alpha$) should yield bound-state representations, which are also short (two-dimensional) and have the same structure outlined above, in~\eqref{eq:repr-bosonic}. Schematically,
\begin{equation}
\rho^\B_\L(|m|,p) \otimes \rho^\B_\L(|m|,p) \supset \rho^\B_\L(2|m|,p) \,, \quad  \rho^\F_\R(-|m|,p) \otimes \rho^\F_\R(-|m|,p) \supset \rho^\F_\R(-2|m|,p) \,, 
\end{equation}
and so on. These two-dimensional representation can be constructed by fusing either two h.w.s.\ or two l.w.s., depending on whether we are dealing with left or right representations,%
\footnote{In particular, we expect bound-state poles in the S-matrix elements of h.w.s.\ in the left representation ($S_{YY}$, $S_{XX}$), and between l.w.s.\ in right ($S_{\bar{Y} \bar{Y}}$, $S_{\bar{X} \bar{X}}$).}
and acting with the supercharges to create the remaining state.
The S-matrix elements featuring poles are 
\begin{equation}
\label{eq:sec_fusion_str_AF_S_connection}
\mathcal{S}_{YY} \,, \qquad
\mathcal{S}_{\bar{Y} \bar{Y}} \,, \qquad 
\mathcal{S}_{XX} \,, \qquad 
\mathcal{S}_{\bar{X} \bar{X}}\,.
\end{equation}
Here and below we use the calligraphic font ($\mathcal{S}$ rather than $S$) to highlight that we are referring to the full S-matrix element, including its dressing factor and CDD factors.

%%%%
\paragraph{Fusion in the mirror model.}

The bound-state condition in mirror kinematics is different from the one of the string kinematics~\cite{Arutyunov:2007tc}.\footnote{Working with Zhukovsky variables the pole condition in the string theory is $x^{+m}_1=x^{-m}_2$, while in the mirror theory we have $x^{-m}_1=x^{+m}_2$.}
Still, from the pole structure for the string model, using unitarity and analitycity, we see that the elements $\mathcal{S}_{YY},\mathcal{S}_{XX}$ and $\mathcal{S}_{\bar{Y}\bar{Y}},\mathcal{S}_{\bar{X}\bar{X}}$ have no poles, though they have a simple zero, and that the scattering of fermions has no poles or zeros. It is instead reasonable to conjecture the existence of mirror bound-state poles in the elements associated with particles related to different spheres:
\begin{equation}
\mathcal{S}_{\psi \chi} \,, \qquad \mathcal{S}_{\bar{\psi} \bar{\chi}}\,.
\end{equation}
For generic values of $\alpha$, these have different masses $|m_1|=\alpha$ and $|m_2|=1-\alpha$. However, even in the special case $\alpha=1/2$ we may distinguish these states by looking at the $\mathfrak{u}(1)$ charges in~\eqref{eq:u1charges}.
Similarly to~\cite{Frolov:2023lwd}, we do not expect the mirror theory to have a well-defined relativistic limit. Taking the relativistic limit of the string model and then applying a mirror transformation (which is trivial for a relativistic model) need not commute with doing the mirror transformation first and then applying the relativistic limit on the mirror theory.  In fact, it is unclear how to take the relativistic limit of the mirror model, since the latter is non-unitary~\cite{Baglioni:2023zsf}.

%%%%%%%%%%%%%%%%%%%%%%%%%
\section{Relativistic limit}
\label{sec:limit}

Let us consider the relativistic of mixed-flux $\AdSSSS$ superstrings along the lines of~\cite{Frolov:2023lwd}.%
\footnote{A similar limit was also previously studied in~\cite{Fontanella:2019ury} for $\AdSST$, though with a different interpretation and particle content.}

\subsection{Limit and representations}

We start by observing that the dispersion relation $H_m(p)$ of~\eqref{eq:reprEigen} has a minimum at $p=-2\pi m/k$. If $m/k\notin\mathbb{Z}$, the minimum is quadratic, and it is possible to expand the $H_m(p)$ around
\begin{equation}
\label{eq:massivelimit}
p=-\frac{2 \pi m}{k}+ \frac{2 \pi}{k} \mu_m \sinh\theta +\mathcal{O}(h^2)\,,\qquad \mu_m=2h\,\left|\sin\frac{m\pi}{k}\right|\,.
\end{equation}
Then, the dispersion relation becomes (at leading order in~$h$) that of a \textit{relativistic} particle of mass~$\mu_m$ and rapidity~$\theta$, $H_m(\theta)=\mu_m\cosh\theta$. It is worth noting that \textit{this expansion does not assume that $k$ or $m$ are integer}, but only that $m/k\notin\mathbb{Z}$. The case $m/k\in\mathbb{Z}$ can also be studied. This was done in~\cite{Frolov:2023lwd} and it yields a relativistic \textit{massless} dispersion. Because we are mostly interested in the bound-state structure of the model, and because the analysis of the massless sector is essentially the same as in~\cite{Frolov:2023lwd}, here we focus on the the massive case of~\eqref{eq:massivelimit} with  $m/k\notin\mathbb{Z}$.

In fact, we are especially interested in the limit of the representations of the lightest modes, which we denote with
\begin{equation}
\rho_\L^\B(\alpha, \theta)\,, \qquad \rho_\R^\F(-\alpha, \theta)\,, \qquad \rho_\L^\B(1-\alpha, \theta)\,, \qquad \rho_\R^\F(-(1-\alpha), \theta)\,,
\end{equation} 
where $\theta$ highlights that the limit has been taken.
Similarly to~\eqref{eq:lrisomorphism}, we have the isomorphisms
\begin{equation}
\rho_\R^\F(-\alpha, \theta) \cong \rho_\L^\B(k-\alpha, \theta)\,,\qquad
\rho_\R^\F(-1+\alpha, \theta) \cong \rho_\L^\B(k-1+\alpha), \theta) \,.
\end{equation}
These formulae can be obtained expanding around the minimum~\eqref{eq:massivelimit} both sides of the isomorphisms in~\eqref{eq:lrisomorphism}, or checked independently after the expansion.
In any case, motivated by these isomorphisms we will restrict to study the representations 
\begin{equation}
\label{eq:light_reps}
\rho_\L^\B(\alpha, \theta)\,, \qquad \rho_\L^\B(k-\alpha, \theta)\,, \qquad \rho_\L^\B(1-\alpha, \theta)\,, \qquad \rho_\L^\B(k-(1-\alpha), \theta)\,,
\end{equation}
which encompass all the lightest modes, as well as (as we shall discuss in a moment) their bound states.%
\footnote{%
From now on, we always work with the left representation and bosonic highest weight states, mostly omitting the labels ``B'' and ``L''.}
As we will comment later, this identification is compatible not only with the symmetries of the model, but with the S-matrix bootstrap (i.e., with the construction of the dressing factors for bound states).

%%%%%%%%%%%%%%%%%%%%%%%%%%
\paragraph{Bound-state masses.}

Let us look at on the fusion properties of the theory and its bound states in the relativistic limit. Recall that the relativistic mass is $2h\,|\sin \pi m/k|$ and, as reviewed in appendix~\ref{app:reviewlimit}, the S-matrix elements similarly depend on the ratio $m/k$, rather than separately on $m$ and~$k$. We encounter two families of bound states:
\begin{enumerate}
    \item Bound state from the first sphere, composed of $Q\in\mathbb{N}$ particles of mass $\alpha={k_2}/{(k_1+k_2)}$. They have a relativistic mass given by
\begin{equation}
\mu=2h\,\left|\sin\frac{\alpha Q\pi}{k}\right|=2h\,\left|\sin\frac{Q \pi}{k_1}\right| \,.
\end{equation}
\item Bound states from the second sphere, composed of $Q\in\mathbb{N}$ particles of mass $1-\alpha={k_1}/{(k_1+k_2)}$. They have a bound-state relativistic mass
\begin{equation}
\mu=2h\,\left|\sin\frac{(1-\alpha) Q\pi}{k}\right|=2h\,\left|\sin\frac{Q \pi}{k_2}\right| \,.
\end{equation}
\end{enumerate}

Note that if $k_1$ and $k_2$ are not co-prime, it is possible to construct bound states which have the same mass though they come from different spheres. It is sufficient to take $Q_1$ particles from the first sphere and $Q_2$ particles from the second sphere such that
\begin{equation}
\label{eq:condition_for_common_bstate}
\frac{Q_1}{k_1} = \frac{Q_2}{k_2}\,.
\end{equation}
It should be stressed that in the original theory these particles, while having the same mass, would be distinguishable by using the $\mathfrak{u}(1)$ charges listed in~\eqref{eq:u1charges}.

%%%%%%%%%%%%%%%%%%%%%%%%%%
\subsection{Bound-state S matrix}
\label{sec:bound_st_Smat}

The general construction of the bound-state S~matrix in the relativistic model was worked out in~\cite{Frolov:2023lwd} starting from~$\AdSST$, and it is summarised in appendix~\ref{app:reviewlimit}. Schematically, the full S-matrix has the form
\begin{equation}
\mathcal{S}_{Q'Q''}(\theta',\theta'') = \Phi_{Q'Q''}(\theta) \, \sigma^{\text{min}}_{Q'Q''}(\theta) \, S_{Q'Q''}(\theta)\,,\qquad
\theta\equiv\theta'-\theta''\,,
\end{equation}
where we singled out: the matrix part of the S~matrix, denoted by~$S_{Q'Q''}$ and canonically normalised so that the scattering of two h.w.s.\ is one; the minimal solution of the crossing equation~$\sigma^{\text{min}}_{Q'Q''}$ (described in appendix~\ref{app:reviewlimit}) which has no poles in the strip $0<\text{Im}[\theta]<\pi$; and a ``CDD''~\cite{Castillejo:1955ed} prefactor~$\Phi_{Q'Q''}(\theta)$ which accounts for the bound-state structure.
This last function is a product of building blocks of the form
\begin{equation}
\label{eq:polefactor}
\left[\frac{m}{k}\right]_\theta= \frac{\sinh \left( \frac{\theta}{2} + \frac{i\pi m}{2k} \right)}{\sinh \left( \frac{\theta}{2} - \frac{i\pi m}{2k}  \right)}\,,
\end{equation}
which satisfy the homogeneous crossing equation\footnote{More specifically, they satisfy the following simple crossing equation
$\left[\frac{m}{k}\right]_\theta \, \left[\frac{k-m}{k}\right]_{\theta+i\pi}=-1$ relating particles of positive energies and momenta with anti-particles with negative energies and momenta.}. 
In the case of~$\AdSST$, $k$ was an integer. Here, we will take instead $k=k_1k_2/(k_1+k_2)$ as in eq.~\eqref{eq:alphakmap}.

When referring to a specific S-matrix element (including the dressing factors), we use the notation $\mathcal{A}_{Q' Q''}$, $\mathcal{B}_{Q' Q''}$, through $\mathcal{F}_{Q' Q''}$ for the full S-matrix elements. In particular, $\mathcal{A}_{Q' Q''}$ corresponds to the scattering of the h.w.s., while $\mathcal{F}_{Q' Q''}$ to the scattering of two l.w.s.; both processes are elastic.
The ``matrix part'' is indicated by $A_{Q' Q''}$ through $F_{Q' Q''}$. In our convention, we normalise
$A^{\alpha\alpha}_{Q' Q''}=1$, and the remaining matrix-part coefficients $B^{\alpha\alpha}_{Q' Q''}(\theta)$, $C^{\alpha\alpha}_{Q' Q''}(\theta)=E^{\alpha\alpha}_{Q' Q''}(\theta)$, $D^{\alpha\alpha}_{Q' Q''}(\theta)$, and $F^{\alpha\alpha}_{Q' Q''}(\theta)$ are fixed by symmetry, see~\eqref{eq:ZFalgebraappendix}.

A minimal set-up for the spectrum involves two families of bound states that should be generated from the fusion of the particles of mass $\alpha$ and $1-\alpha$, separately.%
\footnote{Recall that even in the special case $\alpha=1/2$ the two families can in principle be distinguished by making use of the $\mathfrak{u}(1)$ charges of~\eqref{eq:u1charges}.}
Therefore, we will discuss separately the scattering of bound state coming from the same sphere, and from different spheres.

\paragraph{Scattering of  two bound states from the first sphere.}
A bound state made of $Q'$ particles of mass $\alpha$ appears to scatter with a similar bound state of $Q''$ particles by means of the S-matrix which is normalised as
\begin{equation}
\mathcal{A}^{\alpha \alpha}_{Q' Q''} (\theta) = \Phi^{\alpha\alpha}_{Q' Q''}(\theta) \, \sigma^{\alpha\alpha,\text{min}}_{Q' Q''}(\theta) \, A^{\alpha\alpha}_{Q' Q''}(\theta) \,, \qquad Q' , Q''= 1, \, 2, \, \dots, k_1 - 1 \,.
\end{equation} 
The representations and resulting matrix part of appendix~\ref{app:reviewlimit} apply to the case at hand with minor modifications: because representations only depend on the ration $m/k$ and here
\begin{equation}
    \frac{m}{k}=\frac{Q\alpha}{k}=\frac{Q}{k_1}\,,
    \qquad
    \left[\frac{m}{k}\right]_\theta= \left[\frac{Q}{k_1}\right]_\theta\,,
    \qquad Q=1,\dots, k_1-1\,,
\end{equation}
the S-matrix elements of~\eqref{eq:Smatrixelements-appendix} can be modified to give the $A^{\alpha\alpha}_{Q' Q''}(\theta)$, $B^{\alpha\alpha}_{Q' Q''}(\theta)$, etc., just by replacing $k$ with $k_1$ and $m', m''$ with $Q', Q''$.
Moreover, it is easy to check that a minimal solution of the crossing equations can be found by making a similar substitution in the $\AdSST$ result, namely by setting
\begin{equation}
\sigma^{\alpha\alpha, \text{min}}_{Q' Q''}(\theta)=\frac{R\left(\theta-\frac{i\pi(Q'+Q'')}{k_1}\right)\,R\left(\theta+\frac{i\pi(Q'+Q'')}{k_1}\right)}{R\left(\theta-\frac{i\pi(Q'-Q'')}{k_1}\right) \,R\left(\theta+\frac{i\pi(Q'-Q'')}{k_1}\right)} \,, \qquad Q', Q''=1,  \dots,  k_1-1 \,.
\end{equation}
Finally, the CDD factor which accounts for the bound-state poles is given by
\begin{equation}
    \Phi^{\alpha\alpha}_{Q' Q''}(\theta)=
    \left[\frac{Q'+Q''}{k_1}\right]_\theta \, \left[\frac{Q'+Q''-2}{k_1} \right]^2_\theta \, \dots \, \left[\frac{|Q'-Q''|+2}{k_1} \right]^2_\theta \, \left[\frac{|Q'-Q''|}{k_1} \right]_\theta,
\end{equation}
which again is a simple modification of the results for $\AdSST$, see~\eqref{eq:CDD_proposal_old}.

\paragraph{Scattering of two bound states from the second sphere.}
The discussion of this case is completely analogous to the previous one. We have
\begin{equation}
\mathcal{A}^{1-\alpha, 1-\alpha}_{Q' Q''} (\theta) = \Phi^{1-\alpha,1-\alpha}_{Q' Q''}(\theta) \, \sigma^{1-\alpha,1-\alpha,\text{min}}_{Q' Q''}(\theta) \, A^{1-\alpha,1-\alpha}_{Q' Q''}(\theta) \,,
\end{equation}
where now the various functions involve $k_2$ instead of $k_1$, and $Q' , Q''= 1, \dots, k_2 - 1$.

\paragraph{Scattering of two bound states from either sphere.}
In this case too, representations involved in the scattering follows from those of appendix~\ref{app:reviewlimit}. In this case, it boils down to setting $m'/k=Q'/k_1$ and $m''/k=Q''/k_2$ for the two particles (in the case where the first particle is related to the first sphere), where $Q'=1,\dots, k_1-1$ and $Q''=1,\dots, k_2-1$. By way of example, the resulting S-matrix elements take the form
\begin{equation}
    A^{\alpha,1-\alpha}_{Q' Q''}(\theta)= 1 \, , \qquad B^{\alpha,1-\alpha}_{Q'Q''}(\theta)= \frac{\sinh \Bigl(\frac{\theta}{2} - \frac{i \pi}{2} (\frac{Q'}{k_1}-\frac{Q''}{k_2}) \Bigr)}{\sinh \Bigl(\frac{\theta}{2} + \frac{i \pi}{2} (\frac{Q'}{k_1}+\frac{Q''}{k_2}) \Bigr)} \,, \qquad \dots \,,
\end{equation}
where $A^{\alpha,1-\alpha}_{Q' Q''}= 1$ is the usual normalisation for the matrix part of the S~matrix.
The correct normalisation of the h.s.w.\ scattering is given by
\begin{equation}
\mathcal{A}^{\alpha,1-\alpha}_{Q' Q''} (\theta) = \Phi^{\alpha,1-\alpha}_{Q' Q''}(\theta) \, \sigma^{\alpha,1-\alpha,\text{min}}_{Q' Q''}(\theta) \, A^{\alpha,1-\alpha}_{Q' Q''}(\theta) \,,
\end{equation}
with $Q'=1,\dots,k_1-1$ and $Q''= 1, \dots, k_2 - 1$. 
It is easy to check that the minimal solution of the crossing equations in this case is
\begin{equation}
\label{eq:minimal-mixed-ass}
\sigma^{\alpha,1-\alpha, \text{min}}_{Q' Q''}(\theta) =  \frac{R\left(\theta-i\pi(\frac{Q'}{k_1}+\frac{Q''}{k_2})\right)\,R\left(\theta+i\pi(\frac{Q'}{k_1}+\frac{Q''}{k_2})\right)}{R\left(\theta-i\pi(\frac{Q'}{k_1}-\frac{Q''}{k_2})\right) \,R\left(\theta+i\pi(\frac{Q'}{k_1}-\frac{Q''}{k_2})\right)} \,.
\end{equation}
As for the CDD factor $\Phi^{\alpha,1-\alpha}_{Q' Q''}(\theta)$, we do not expect any bound state between particle related to different spheres, and therefore we must choose
\begin{equation}
\label{eq:trivialcdd}
    \Phi^{\alpha,1-\alpha}_{Q' Q''}(\theta)=1\,.
\end{equation}
While it is clear that the choice~\eqref{eq:trivialcdd} is necessary, it is possible to check that it is sufficient: that is to say, eq.~\eqref{eq:minimal-mixed-ass} has no poles for any $Q'=1,\dots,k_1-1$ and $Q''= 1, \dots, k_2 - 1$ in the physical strip $0<\text{Im}[\theta]<\pi$.
Finally, the S~matrix $\mathcal{S}^{1-\alpha, \alpha}_{Q', Q''}(\theta)$ is connected to the one above by braiding unitarity, and we have in particular
\begin{equation}
    \mathcal{A}^{1-\alpha,\alpha}_{Q' Q''}(\theta)=
    \left(\mathcal{A}^{\alpha,1-\alpha}_{Q'' Q'}(-\theta)\right)^{-1},\qquad Q'=1,\dots,k_2-1\,,\quad Q''=1,\dots,k_1-1\,.
\end{equation}

%%%%%%%%%%%%
\paragraph{Bound-state identification.}
Because of the isomorphism~\eqref{eq:lrisomorphism}, ``right'' representations with negative $m=-|m|$ (with $0<|m|<k$) can be identified with left representations of mass $k-|m|$. Before the relativistic limit, this identification is not possible for the full S matrix, due to monodromy factors, as worked out for $\AdSST$~\cite{Frolov:2025uwz}. As it turns out, such factor become trivial in the relativistic limit. This suggests that in the relativistic limit we may identify the distinguished S-matrix elements of~\eqref{eq:sec_fusion_str_AF_S_connection} with
\begin{equation}
\label{eq:smatrix-id}
\mathcal{A}^{\alpha\alpha}_{1,1}\equiv \mathcal{S}_{YY} \,, \quad 
\mathcal{A}^{1-\alpha, 1-\alpha}_{1,1}\equiv \mathcal{S}_{XX}\,,\quad
\mathcal{F}^{\alpha\alpha}_{k_1-1, k_1-1}\equiv S_{\bar{Y} \bar{Y}} \,, \quad \mathcal{F}^{1-\alpha,1-\alpha}_{k_2-1, k_2-1}\equiv S_{\bar{X} \bar{X}}\,,
\end{equation}
similarly to what done for $\AdSST$.
This choice is consistent with our decision to restrict from the get-go to representations with a bosonic h.w.s., see appendix~\ref{app:Low_energy_limit_h_finite}.%
\footnote{One may wonder why, if we only work with representations having \textit{bosonic highest-weight states}, we identify the boson-boson scattering $S_{\bar{Y} \bar{Y}}$ with the scattering of \textit{lowest} weight states~$\mathcal{F}^{\alpha\alpha}_{k_1-1, k_1-1}$ (and likewise for $S_{\bar{X} \bar{X}}$). The reason is that the coproduct in the supercharges~\eqref{eq:limit_of_supercharge_Qm1m2} involves a factor of the type $e^{i\pi m/k}=e^{i\pi Q/k_1}$ so that shifting $Q\to Q+k$ results in an additional sign, which ``swaps'' the statistics~\cite{Frolov:2023lwd}.}
In principle, we could have introduced representations with fermionic h.w.s.\ too, like it happens in the full model, and carried out the bootstrap for both. In that case, we would have naturally identified instead $S_{\bar{Y} \bar{Y}}=\mathcal{F}^{-\alpha,-\alpha}_{1, 1}$, where this is the S-matrix element that scatters two (bosonic) l.w.s.\ belonging the limit  of the ``right'' representation (whose h.w.s.\ is fermionic). This would have lead to the very same result as~\eqref{eq:smatrix-id}, consistently with the observation that the monodromy factors of~\cite{Frolov:2025uwz} trivialise in the relativistic limit.
A final consistency check is the observation that the S-matrix elements $\mathcal{A}^{\alpha\alpha}_{1,1}$ and  $\mathcal{F}^{\alpha\alpha}_{k_1-1, k_1-1}$ have the correct bound-state poles to ensure the expected fusion (while $\mathcal{F}^{\alpha\alpha}_{1, 1}$  and $\mathcal{A}^{\alpha\alpha}_{k_1-1,k_1-1}$ do not have bound-state poles). The same holds for the second sphere.

\paragraph{Accidentally equal masses.}
As we mentioned, if $k_1$ and $k_2$ are not co-prime it is possible to construct bound states which have the same mass though they come from different spheres, simply by setting $Q_1/k_1=Q_2/k_2$.
It is easy to verify that the fused S-matrix elements are different, 
\begin{equation}
\mathcal{S}^{\alpha \alpha}_{Q_1 Q_1} \ne
\mathcal{S}^{1-\alpha, 1-\alpha}_{Q_2 Q_2} \ne
\mathcal{S}^{\alpha, 1-\alpha}_{Q_1 Q_2} \,.
\end{equation}
We then need to separate the bound states into two sectors, keeping track of whether they are made of particles of types $\alpha$ or $1-\alpha$. This is consistent at the level of fusion, as it is shown in some detail in appendix~\ref{sec:app_separate_sectors} on an example. With this picture, there are pairs of representations sharing the same mass but having different S-matrix elements. This should not be a surprise since these representations have bosons living on different three-spheres.  This picture agrees with the model described in~\cite{Frolov:2025syj}.

\paragraph{Summary of the results.}
For convenience, we spell out the explicit value of the S-matrix elements corresponding to fundamental particles of the original model. For ``left'' particles we have, after the relativistic limit
\begin{equation}
\begin{split}
\mathcal{S}_{Y Y}(\theta)&=\mathcal{A}^{\alpha \alpha}_{1 ,1} (\theta) = \frac{\sinh \left( \frac{\theta}{2} + \frac{i \pi}{k_1} \right)}{\sinh \left( \frac{\theta}{2} - \frac{i \pi}{k_1} \right)}\frac{R\left(\theta-\frac{2i\pi}{k_1}\right)\,R\left(\theta+\frac{2i\pi}{k_1} \right)}{R(\theta)^2} \,, \\
\mathcal{S}_{X X}(\theta)&=\mathcal{A}^{1-\alpha,1- \alpha}_{1 1} (\theta) = \frac{\sinh \left( \frac{\theta}{2} + \frac{i \pi}{k_2} \right)}{\sinh \left( \frac{\theta}{2} - \frac{i \pi}{k_2} \right)}\frac{R\left(\theta-\frac{2i\pi}{k_2}\right)\,R\left(\theta+\frac{2i\pi}{k_2} \right)}{R(\theta)^2} \,,\\
\mathcal{S}_{Y X}(\theta)&=\mathcal{A}^{\alpha,1- \alpha}_{1, 1} (\theta) = \frac{R\left(\theta-i \pi (\frac{1}{k_1} + \frac{1}{k_2})\right)\,R\left(\theta + i \pi (\frac{1}{k_1} + \frac{1}{k_2}) \right)}{R \left(\theta-i \pi (\frac{1}{k_1} - \frac{1}{k_2}) \right) R \left(\theta+i \pi (\frac{1}{k_1} - \frac{1}{k_2}) \right)} \,.
\end{split}
\end{equation}
The  ``right'' particles of the original model are related to bound states, so that
\begin{equation}
\label{eq:rel_limit_right_particles}
\begin{split}
\mathcal{S}_{\bar{Y} \bar{Y}}(\theta)&=\mathcal{F}^{\alpha \alpha}_{k_1-1 , k_1-1} (\theta)=\frac{\sinh \left( \frac{\theta}{2} + \frac{i \pi}{k_1} \right)}{\sinh \left( \frac{\theta}{2} - \frac{i \pi}{k_1} \right)}\frac{R\left(\theta-\frac{2i\pi}{k_1}\right)\,R\left(\theta+\frac{2i\pi}{k_1} \right)}{R(\theta)^2} \,, \\
\mathcal{S}_{\bar{X} \bar{X}}(\theta)&=\mathcal{F}^{1-\alpha,1- \alpha}_{k_2-1 , k_2-1} (\theta) = \frac{\sinh \left( \frac{\theta}{2} + \frac{i \pi}{k_2} \right)}{\sinh \left( \frac{\theta}{2} - \frac{i \pi}{k_2} \right)}\frac{R\left(\theta-\frac{2i\pi}{k_2}\right)\,R\left(\theta+\frac{2i\pi}{k_2} \right)}{R(\theta)^2} \,,\\
\mathcal{S}_{\bar{Y} \bar{X}}(\theta)&=\mathcal{F}^{\alpha,1- \alpha}_{k_1-1, k_2-1} (\theta) = \frac{R\left(\theta-i \pi (\frac{1}{k_1} + \frac{1}{k_2})\right)\,R\left(\theta + i \pi (\frac{1}{k_1} + \frac{1}{k_2}) \right)}{R \left(\theta-i \pi (\frac{1}{k_1} - \frac{1}{k_2}) \right) R \left(\theta+i \pi (\frac{1}{k_1} - \frac{1}{k_2}) \right)} \,.
\end{split}
\end{equation}

\subsection{Alternative S~matrix from a fictitious fundamental particle}
\label{sec:fictitious_min_part}

If we forget about the $\mathfrak{u}(1)$ charges listed in~\eqref{eq:u1charges}, and give up the possibility of distinguishing which particle comes from which sphere --- other than by looking at its mass --- we can describe all particles of the theory in terms of a single excitations of ``minimal mass'', as suggested in the conclusions of~\cite{Frolov:2025syj}. Because this particle does not appear in the spectrum but it is a tool to construct it, we call it \textit{fictitious}. It is not immediately clear what should be this ``minimal mass''. The two natural options are
\begin{equation}
\label{eq:kzero}
    \mu_0 = 2h\,\left|\sin\frac{\pi}{k_0}\right|\,,\qquad
    k_0=\begin{cases}
        k_1\,k_2&\text{(option A),}\\
        \text{mcm}(k_1,k_2)&\text{(option B),}
    \end{cases}
\end{equation}
where ``mcm'' stands for minimum common multiple.
Option A would be consistent that the view that the model depends on $k_1$ and $k_2$, rather than on $\alpha$ itself, and that it makes sense to consider ``the lightest possible fictitious mass'' that can be constructed out of $k_1$ and $k_2$ in a natural way. 
Option B instead chooses the heaviest possible fictitious mass.%
\footnote{This is illustrated by the case $\alpha=1/2$, that is $k_1=k_2=2k$: with option~A we would take $k_0=k_1k_2=4k^2$ whereas with option B we would take $k_0=k_1=k_2=2k$.}
We will see below that these two options lead to different dressing factors.
In any case, it is clear that in either option we may obtain the original light particles by choosing the bound-state number so that having $Q=k_0/k_1$ or $Q=k_0/k_2$ (in option~A we would simply pick $Q=k_2$ or $Q=k_1$, respectively).
For this reason, below we will just work implicitly in terms of~$k_0$ and only at the end comment on the difference between case A and~B.

\paragraph{Fusion.}
The construction of the fused S-matrix elements follows the same logic as in the previous sub-section. In particular we may set
\begin{equation}
\mathcal{A}_{Q'Q''} (\theta) = \Phi_{Q'Q''}(\theta) \, \sigma^{\text{min}}_{Q'Q''}(\theta) \, A_{Q'Q''}(\theta) \,,\qquad Q',Q''=1,\dots,k_0-1\,,
\end{equation}
where the various pieces of the S-matrix element take the form in appendix~\ref{app:reviewlimit} up to the replacement $k\to k_0$. In particular, recall that in our conventions $A_{Q'Q''}=1$ and that for the minimal dressing factor we have
\begin{equation}
\sigma^{\text{min}}_{Q'Q''}(\theta)=\frac{R\left(\theta-\frac{i\pi(Q'+Q'')}{k_0}\right)\,R\left(\theta+\frac{i\pi(Q'+Q'')}{k_0}\right)}{R\left(\theta-\frac{i\pi(Q'-Q'')}{k_0}\right) \,R\left(\theta+\frac{i\pi(Q'-Q'')}{k_0}\right)} \,, 
\end{equation}
and the CDD term is
\begin{equation}
\Phi_{Q'Q''}(\theta)= \left[\frac{Q'+Q''}{k_0}\right]_\theta \,
\left[\frac{Q'+Q''-2}{k_0}\right]_\theta^2\,
\cdots
\left[\frac{|Q'-Q''|+2}{k_0}\right]_\theta^2
\left[\frac{|Q'-Q''|}{k_0}\right]_\theta\,.
\end{equation}

\paragraph{Identifying the particles of the original model.}
Original excitations of the model should be identified with $Q$-particle bound states so that $Q/k_0=\alpha$ or $Q/k_0=1-\alpha$. 
For instance, for the scattering of two $m=\alpha$ particles ($Q=k_0/k_1$) we find the S-matrix element
\begin{equation}
\mathcal{S}_{YY}=\mathcal{A}_{Q Q}(\theta) = 
\left[\frac{2Q}{k_0}\right]_\theta\,
\left[\frac{2Q-2}{k_0}\right]_\theta^2\,\cdots
\left[\frac{2}{k_0}\right]_\theta^2\,
\left[0\right]_\theta \, \sigma^{\text{min}}_{Q Q} (\theta) \,.
\end{equation}
The building block
\begin{equation}
\left[\frac{2Q}{k_0}\right]_\theta= \frac{\sinh \left( \frac{\theta}{2} + \frac{i \pi Q}{k_0} \right)}{\sinh \left( \frac{\theta}{2} - \frac{i \pi Q}{k_0} \right)}= \frac{\sinh \left( \frac{\theta}{2} + \frac{i \pi}{k_1} \right)}{\sinh \left( \frac{\theta}{2} - \frac{i \pi}{k_1} \right)}
\end{equation}
has a simple pole at $\theta=2\pi i /k_1$, or equivalently $\theta=2\pi i\alpha/k$.
This is consistent with the existence of bound states between particles of mass $\alpha$, as we want. A similar pole appears in the S-matrix element $\mathcal{F}_{k_0-Q, k_0-Q}(\theta)=\mathcal{S}_{\bar{Y}\bar{Y}}$, which is again expected. In addition to these bound states, we observe many second-order singularities of Coleman--Thun type~\cite{Coleman:1978kk}, coming from the fact that we are introducing particles of mass smaller than $\alpha$. As a consequence, even if we restrict to fundamental particle of the original model (the particles of mass $\alpha$, $1-\alpha$ and their bound states), the S-matrix elements would show the signature of the existence of lighter particles in the form of double poles.

\paragraph{Bound states between different spheres.}
In this picture, scattering of excitations related to either sphere can lead to bound states too. Considering for instance the S-matrix element scattering h.w.s.\ particles of mass $\alpha$ and $1-\alpha$, we find a bound-state pole for a particle of mass~$1$. To see this, choose $Q_1$ and $Q_2$ such that $Q_1=k_0/k_1$ and $Q_2=k_0/k_2$ so that
\begin{equation}
\mathcal{A}_{Q_1 Q_2}(\theta) =
\left[\frac{Q_1+Q_2}{k_0}\right]_\theta 
\left[\frac{Q_1+Q_2-2}{k_0}\right]_\theta^2\cdots
\left[\frac{|Q_1-Q_2|+2}{k_0}\right]_\theta^2 
\left[\frac{|Q_1-Q_2|}{k_0}\right]_\theta
 \sigma^{\text{min}}_{Q_1 Q_2} (\theta).
\end{equation}
The building block $[(Q_1+Q_2)/k_0]_\theta$ has a simple pole at
\begin{equation}
\label{eq:mixed-sphere-pole}
\theta=i \pi \frac{Q_1+Q_2}{k_0} =i\pi\left(\frac{1}{k_1}+\frac{1}{k_2}\right)=\frac{i\pi}{k}\,,
\end{equation}
associated with the scattering of a particle of mass $1$, while (taking into account the contribution of $F_{Q_1Q_2}$) the full element $\mathcal{F}_{Q_1 Q_2}$ is regular. 
It is tempting to identify this bound-state with the relativistic limit of the $\text{AdS}_{3}$ excitations, but the resulting bound-state representation has a bosonic h.w.s., rather than a fermionic one as expected from~\eqref{eq:adsboson}. 
Another concern is the pole structure for this process:  if the pole~\eqref{eq:mixed-sphere-pole} existed in $\mathcal{S}_{XY}$ in the string model \textit{before the relativistic limit}, then it should be mapped to a zero of the mirror model. As a result, the mirror model could not have any bound states at all between particles of mass $\alpha$ and $1-\alpha$.
This would be in contrast with the pole structure and string hypothesis of~\cite{Frolov:2025syj}.

%%%%%%%%%%%%%%%%%%%%%%%%%%
\subsection{Comparison with Frolov-Sfondrini phases}
\label{sec:comparison}

Let us compare the dressing factors proposed here with the relativistic limit of the results in~\cite{Frolov:2025syj}, where the authors advanced the following proposals for particles of equal and different masses
\begin{equation}
\label{eq:SXXm1m2}
\begin{split}
\mathcal{S}^{m m}_{\bar{\cX} \bar{\cX}}(u_1,u_2)=&+ H^{mm}_{\bar{\cX} \bar{\cX}}  (u_1,u_2)\, {{\tx_{\R1}^{+m}}\ov {\tx_{\R1}^{-m}}}\, {{\tx_{\R2}^{-m}}\ov {\tx_{\R2}^{+m}}}\,\left({ \tx^{-m}_{\R1}- \tx^{+m}_{\R2}\ov \tx^{+m}_{\R1}- \tx^{-m}_{\R2}}\right)^2
    \frac{u_1-u_2 + {2im\ov h}}{u_1-u_2 - {2im\ov h}}\\
    &\quad\times
    \frac{R(\tg^{+m+m}_{\R\R}) R(\tg^{-m-m}_{\R\R})}{R(\tg^{+m-m}_{\R\R}) R(\tg^{-m+m}_{\R\R}) }
    \left(\frac{\Sigma^{\bes}_{\R\R}(\tx^{\pm m}_{\R1}, \tx^{\pm m}_{\R2})}{\Sigma^{\hl}_{\R\R}(\tx^{\pm m}_{\R1}, \tx^{\pm m}_{\R2})}\right)^{-2}\,,\\
\mathcal{S}^{m_1m_2}_{\bar{\cX} \bar{\cX}} (u_1,u_2)=&+H^{m_1m_2}_{\bar{\cX} \bar{\cX}} (u_1,u_2)\,\frac{R(\tg^{+m_1+m_2}_{\R\R}) R(\tg^{-m_1-m_2}_{\R\R})}{R(\tg^{+m_1-m_2}_{\R\R}) R(\tg^{-m_1+m_2}_{\R\R}) } \,, \qquad m_1 \ne m_2 \,.
    \end{split}
\end{equation}
We refer to~\cite{Frolov:2025syj} for the notation and definition (which follows the standard conventions for $\text{AdS}_3$ integrability).
The formulae above are valid for ``right'' particles of type $\alpha$ and $1-\alpha$.
The function $H^{m_1m_2}$ is a simple CDD factor given by
\begin{equation}
H^{m_1m_2}_{\cX\cX}  (u_1,u_2)= e^{  -\frac{i}{2}\frac{1-m_2}{m_1}(\tp_1\tE_2 -\tp_2\tE_1)} \qquad \text{(in the mirror region)} \,,
\end{equation}
and $\bar{\cX}=\bar{Y}, \, \bar{X}$.
One should first continue these expression to the string region and then compute the relativistic limit of the matrix elements above. The procedure is described in detail in appendix~K of~\cite{Frolov:2025uwz}, where it was carried out for $\AdSST$. 
The factor $H^{m_1m_2}_{\cX\cX}$, after continuation to the string region, becomes simply
\begin{equation}
H^{m_1m_2}_{\cX\cX}  (u_1,u_2)= e^{  \frac{i}{2}\frac{1-m_2}{m_1}(E_1p_2 -E_2 p_1)}=1 + \mathcal{O}(h) \,.
\end{equation}
Indeed, since in the relativistic limit the energies are of order $h$, then at the leading order this term is one.
Following~\cite{Frolov:2025uwz}, we can continue the remaining terms to the string region and obtain their relativistic limits. It turns out that in the limit we have
\begin{equation}
\frac{R(\gamma^{+m_1+m_2}_{\R\R}) R(\gamma^{-m_1-m_2}_{\R\R})}{R(\gamma^{+m_1-m_2}_{\R\R}) R(\gamma^{-m_1+m_2}_{\R\R}) } =  \frac{R(\theta+\frac{i \pi}{k} (m_1+m_2)) R(\theta-\frac{i \pi}{k} (m_1+m_2) )}{R(\theta+\frac{i \pi}{k} (m_1-m_2)) R(\theta-\frac{i \pi}{k} (m_1-m_2))}+\mathcal{O}(h) \,.
\end{equation}
Using the $\kappa$-deformed Zhukovsky map~\cite{Stepanchuk:2014kza,Frolov:2023lwd}
\begin{equation}
u_\R(x)=x + \frac{1}{x} + \frac{k}{\pi h} \ln x
\end{equation}
we obtain 
\begin{equation}
u_\R(x^{\pm m}_{\R}) = -\frac{k}{\pi h} \theta \pm i \frac{m}{h} +\mathcal{O}(h^0)\,.
\end{equation}
If we define 
\begin{equation}
h_m\equiv\frac{h}{m}\,,
\end{equation}
then for $m=\alpha$ and $m=1-\alpha$ we have respectively
\begin{equation}
\begin{split}
u_\R(x^{\pm \alpha}_{\R}) = -\frac{k_1}{\pi h_\alpha} \theta \pm \frac{i}{h_\alpha}+\mathcal{O}(h^0) \,, \qquad u_\R(x^{\pm (1-\alpha)}_{\R}) =-\frac{k_2}{\pi h_{1-\alpha}} \theta \pm \frac{i}{h_{1-\alpha}}+\mathcal{O}(h^0) \,.
\end{split}
\end{equation}
The modified BES and Hern\'andez--L\'opez~\cite{Hernandez:2006tk} (HL) dressing factors introduced in~\cite{Frolov:2025syj} are obtained by simply replacing $h \to h_m$ in the BES and HL kernels. As a consequence of this fact one obtains the limit of the even part of the dressing factor by replacying $k$ with either $k_1$ or $k_2$, in the results of~\cite{Frolov:2025uwz}. We have
\begin{equation}
\label{eq:rel_lim_u_BES_HL}
\begin{split}
\frac{u_1-u_2 + {2i \alpha \ov h}}{u_1-u_2 - {2i \alpha \ov h}} \left(\frac{\Sigma^{\bes}_{\R\R}(x^{\pm \alpha}_{\R1}, x^{\pm \alpha}_{\R2})}{\Sigma^{\hl}_{\R\R}(x^{\pm \alpha}_{\R1}, x^{\pm \alpha}_{\R2})}\right)^{-2} &= \frac{\sinh \left( \frac{\theta}{2} - \frac{i \pi}{k_1} \right)}{\sinh \left( \frac{\theta}{2} + \frac{i \pi}{k_1} \right)}+\mathcal{O}(h) ,\\
\frac{u_1-u_2 + {2i (1-\alpha) \ov h}}{u_1-u_2 - {2i (1-\alpha) \ov h}} \left(\frac{\Sigma^{\bes}_{\R\R}(x^{\pm (1-\alpha)}_{\R1}, x^{\pm (1-\alpha)}_{\R2})}{\Sigma^{\hl}_{\R\R}(x^{\pm (1-\alpha)}_{\R1}, x^{\pm (1-\alpha)}_{\R2})}\right)^{-2} &= \frac{\sinh \left( \frac{\theta}{2} - \frac{i \pi}{k_2} \right)}{\sinh \left( \frac{\theta}{2} + \frac{i \pi}{k_2} \right)} +\mathcal{O}(h) .
\end{split}
\end{equation}
From this we find that the S-matrix elements proposed in~\cite{Frolov:2025syj} have the following relativistic limit as $h\to0$:
\begin{equation}
\label{eq:SXXm1m2_2}
\begin{split}
\mathcal{S}_{\bar{Y} \bar{Y}}(u_1,u_2)= & \frac{\sinh \left( \frac{\theta}{2} + \frac{i \pi}{k_1} \right)}{\sinh \left( \frac{\theta}{2} - \frac{i \pi}{k_1} \right)}
    \frac{R \left(\theta+\frac{2i \pi}{k_1} \right) R \left( \theta-\frac{2i \pi}{k_1} \right) }{R(\theta)^2}
     ,\\
\mathcal{S}_{\bar{X} \bar{X}}(u_1,u_2)= & \frac{\sinh \left( \frac{\theta}{2} + \frac{i \pi}{k_2} \right)}{\sinh \left( \frac{\theta}{2} - \frac{i \pi}{k_2} \right)}
    \frac{R \left( \theta+\frac{2i \pi}{k_2} \right) R \left(\theta-\frac{2i \pi}{k_2} \right)}{R(\theta)^2}
     \,,\\
S_{\bar{Y} \bar{X}} (u_1,u_2) =&\frac{R \left(\theta+i \pi (\frac{1}{k_1}+\frac{1}{k_2}) \right) R \left( \theta-i \pi (\frac{1}{k_1}+\frac{1}{k_2}) \right)}{R \left( \theta+i \pi (\frac{1}{k_1}-\frac{1}{k_2}) \right) R \left(\theta-i \pi (\frac{1}{k_1}-\frac{1}{k_2}) \right) } \,,
    \end{split}
\end{equation}
in perfect agreement with the result of our relativisitc bootstrap procedure~\eqref{eq:rel_limit_right_particles}.

In appendix D of~\cite{Frolov:2025syj}, two other alternative solutions for the dressing factors were proposed in the special case $\alpha=1/2$ (which is $k_1=k_2$). 
These solutions modify the combination of certain S~matrix elements by introducing or removing a term of the type\footnote{Notice that for $\alpha=1/2$ we have $\alpha=1-\alpha$ and we can set the value of all masses to be $m_1=m_2=m$.}
\begin{equation}
\label{eq:comb_u_CDD_BES_HL}
\frac{x^{-m}_{\R1} - x^{+m}_{\R2}}{x^{+m}_{\R1} - x^{-m}_{\R2}} \, \frac{u_1-u_2 + {2i m \ov h}}{u_1-u_2 - {2i m \ov h}} \left(\frac{\Sigma^{\bes}_{\R\R}(x^{\pm m}_{\R1}, x^{\pm m}_{\R2})}{\Sigma^{\hl}_{\R\R} (x^{\pm m}_{\R1}, x^{\pm m}_{\R2}) }\right)^{-2} \,.
\end{equation}
Although these solutions are compatible with all the properties of the model, they present some issues with fusion in the mirror kinematics of the non-relativistic model~\cite{Frolov:2025syj}; moreover, they do not admit an obvious generalisation to values of $\alpha\neq1/2$.
As it turns out, the combination of terms in~\eqref{eq:comb_u_CDD_BES_HL} is equal to one in the relativistic limit. Therefore we are unable to establish which proposal of~\cite{Frolov:2025syj} is correct based on the results after the limit.
This is not entirely surprising, because in the relativistic limit the only ambiguity is the location of the bound-state poles, and the expression~\eqref{eq:comb_u_CDD_BES_HL} is regular at $u_1-u_2=\pm 2i m / h$.

\section{Conclusions}
\label{sec:conclusions}

In this paper we have studied the relativistic limit of strings on $\AdSSSS$ supported by mixed RR and NSNS fluxes. The analysis closely resembles that of~\cite{Frolov:2023lwd} and we find a relativistic S-matrix which is closely related to that of Fendley and Intriligator~\cite{Fendley:1992dm}. Still, the presence of two three-spheres in the geometry (or in terms of the worldsheet dynamics, of two integers $k_1$ and $k_2$ related to the three-sphere fluxes) provides a  novel and interesting dynamics, allowing for different bound-state structures --- recall that the dynamics of the $\AdSST$ relativistic limit depended on a single integer~$k$.

The most natural way to realise the S-matrix bootstrap is to assume that particles related to one sphere may make bound-states with each other. Starting from the lightest fundamental excitations, which have mass $\mu=2h\sin(\pi/k_1)$ and $\mu=2h\sin(\pi/k_2)$ in the relativistic limit, we find two families of particles, of mass
\begin{equation}
\label{eq:boundstatespect}
    \mu=2h\sin\frac{Q_1\pi}{k_1}\,,\quad Q_1=1,\dots, k_1-1\,,\qquad
    \mu=2h\sin\frac{Q_2\pi}{k_2}\,,\quad Q_2=1,\dots, k_2-1\,.
\end{equation}
The resulting S-matrix elements can be constructed explicitly, and are closely related to those of~\cite{Fendley:1992dm}. Interestingly, even when $Q_1/k_1=Q_2/k_2$, we can distinguish particles related to two different spheres, and the S-matrices look different. This is consistent with the fact that, before taking the relativistic limit, there are $\mathfrak{u}(1)$ charges in the model which distinguish excitations related to different spheres, cf.~\eqref{eq:u1charges}.
Another way to work out the S-matrix bootstrap, which was mentioned in the conclusion of~\cite{Frolov:2025syj}, is to introduce a light ``fictitious'' particle, with mass $\mu_0=2h\sin(\pi/k_0)$ such that $k_0$ is a common multiple of both $k_1$ and~$k_2$.
In this picture, what we had considered as fundamental particles are now bound-states of $Q=k_0/k_1$ or $k_0/k_2$ particles. The suggestion of~\cite{Frolov:2025syj} was to use the mass-$\mu_0$ particle as a building block of the S~matrix, but restrict the spectrum of physical particles to those emerging as bound states of the ``original'' particles of mass $\mu=2h\sin(\pi/k_1)$ and $\mu=2h\sin(\pi/k_2)$.
In this picture, the fusion changes quite drastically and the S-matrix is qualitatively different from that of the previous case.
The biggest qualitative differences are that now all bound states which have the same mass are to be considered equivalent; in particular, this means that we can construct bound states from excitations from different three-spheres, something that seems not to be the case in the original model before the limit. Moreover, even if we try to use the fictitious light particle as a mere tool to construct the S-matrix, we can find an indication of its existence due to Coleman--Thun poles within other S-matrix elements --- in other words, the S-matrix ``knows'' about the mass $\mu_0$ particle even if we try to restrict the scattering to heavier particles only.

We should also point out that in our analysis we focused on massive excitations. Massless particles decouple from the massive ones in the limit; the same happens for massless particles of different chiralities.
This is very similar to the case of $\AdSST$~\cite{Frolov:2023lwd} and indeed the description of the massless case follows from that paper,%
\footnote{In particular, the same-chirality massless scattering processes follow from eqs.~(C.12) and (C.28) of~\cite{Frolov:2023lwd} by setting the parameter $\alpha$ in those equations (unrelated to ``our'' $\alpha$ in this paper) to $\alpha=\pi$.
}
and for this reason we have not focused much on it here.
We have also not focused much on the ``heavy'' mode related to $\text{AdS}_3$ bosons because there is reason to believe~\cite{Sundin:2012gc} that this is a composite mode, similar to what happens for $\text{AdS}_4\times \mathbb{C}\text{P}^3$ strings~\cite{Zarembo:2009au}. It is worth noting that the $\text{AdS}_3$ mode is not generated by fusion of the sphere modes after the relativistic limit.%
\footnote{In the picture with a ``fictitious'' light particle of mass $\mu_0$, a mode with same mass as the $\text{AdS}_3$ is generated, but it has the wrong statistics to be identified with those excitations, see the discussion around~\eqref{eq:mixed-sphere-pole}.}

This relativistic model is interesting in and of itself, and there are some natural questions that may be worth exploring. 
In analogy to what was done in~\cite{Kervyn:2025ogr}, one may derive its TBA equations and investigate the Y system. Along the lines of~\cite{Torrielli:2023uam}, it may be interesting to investigate the form-factor program  for this model. It may also be interesting to consider special cases, where $k_1=1$, $k_2=1$, or both (in this case, only massless modes survive) or where $k_1=2$, $k_2=2$, or both (in this case, left- and right- excitations are one and the same). This is also similar to $\AdSST$, though of course the structure now is richer due to the presence of two parameters.

Our main motivation for studying this model was to put to the test the recent proposal for the dressing factors of~\cite{Frolov:2025syj}. We find that the natural prescription whereby we build all bound states~\eqref{eq:boundstatespect} from the spheres' excitations yields an S matrix which matches perfectly with the limit of that of~\cite{Frolov:2025syj}. Conversely, the choice where excitations are built out of a fictitious particle of mass $\mu_0$ cannot be reconciled with the proposal of~\cite{Frolov:2025syj}.
It should be noted that while~\cite{Frolov:2025syj} put out a ``main proposal'' for the dressing factors, valid for any~$\alpha$, they also pointed out that more solutions can be constructed for $\alpha=1/2$, though they have problematic fusion properties in the full non-relativistic theory.
Perhaps disappointingly, consistence with the relativistic limit is not enough to rule out those additional solutions. The reason is essentially that they have the same pole structure of the main proposal of~\cite{Frolov:2025syj}, and more nuanced features of the dressing factors are washed away by the relativistic limit.
A proposal for the $\AdSSSS$ dressing factors was also implicitly put forward in the quantum spectral curve construction of~\cite{Cavaglia:2025icd, Chernikov:2025jko}, valid for the case of pure-RR flux and $\alpha=1/2$. In particular, the authors of~\cite{Cavaglia:2025icd} endeavoured to extract the form of the dressing factors from the QSC equations. Unfortunately, a comparison with the results of~\cite{Cavaglia:2025icd, Chernikov:2025jko} is not possible, and would not be so even if a completely explicit expression for the dressing factors had been extracted from the QSC. This relativistic limit crucially hinges on the \textit{mixed-flux} kinematics, while the QSC is currently known for RR-flux models only (even in the case of $\AdSST$). In any case, it currently appears that the QSC cannot reconcile the standard notions of braiding unitarity and crossing symmetry for the dressing factors. This is not the case for our relativistic model, which is perfectly compatible with both. It seems therefore that more work and insight is needed to reconcile the current understanding of the $\AdSSSS$ QSC with the standard axioms of the S-matrix bootstrap on which we relied here.

\section*{Acknowledgments}
We are grateful to Sergey Frolov for numerous useful discussions and collaboration on related work. 
We thank the organisers of the \textit{Integrability, Dualities and Deformations 2025} Workshop at NORDITA, Stockholm for hospitality and for the stimulating environment which contributed to this work.
A.S.\ was supported in part by the CARIPARO Foundation Grant under grant n.~68079.
D.P.\ acknowledges support from the Deutsche Forschungsgemeinschaft (DFG, German Research Foundation) -- SFB-Gesch\"aftszeichen 1624 -- Projektnummer 506632645. 

\noindent

\vskip 10pt

\appendix

\section{Review of the relativistic limit of $\AdSST$}
\label{app:reviewlimit}

In this appendix, we review the relativistic limit of~\cite{Frolov:2023lwd} for the two-particle representations and the S~matrix.

%%%%%%%%%%%%
\subsection{Symmetries and representations}
\label{app:Low_energy_limit_h_finite}
The symmetry algebra for lightcone gauge-fixed  mixed-flux $\AdSST$ superstrings consists of eight supercharges --- twice as many as $\AdSSSS$ --- and it takes the form
\begin{equation}
\label{eq:lcalgebracommT4}
\begin{aligned}
    &\{ \mathbf{Q}^a,  \mathbf{S}_b \}= \frac{\delta^a_b}{2} (\mathbf{H}+\mathbf{M}),\qquad &&\{ \mathbf{Q}^a,  \mathbf{\tilde{Q}}_b \}= \delta^a_b\,\mathbf{C} \; ,\\
    &\{ \mathbf{\tilde{Q}}_a,  \mathbf{\tilde{S}}^b \}= \frac{\delta_a^b}{2} (\mathbf{H}-\mathbf{M}),\qquad &&\{ \mathbf{S}_a,  \mathbf{\tilde{S}}^b \}= \delta_a^b\,\mathbf{\bar{C}} ,
\end{aligned}
\end{equation}
where $a,b=1,2$ are indices of an outer automorphism usually called $\mathfrak{su}(2)_\bullet$~\cite{Sfondrini:2014via}. Short representations are four-dimensional but, as it turns out, they can be constructed by taking tensor product of two-dimensional short representations of the $\AdSSSS$ algebra~\eqref{eq:lcalgebracomm}. Moreover, the eigenvalues of the central charges for $\AdSST$ are the same as in~\eqref{eq:reprEigen} up to requiring $m$ and $k$ to be integers. For this reason, the study of the relativistic limit for $\AdSST$ is closely related to that of $\AdSSSS$ which we outlined around eq.~\eqref{eq:massivelimit}. After the limit, the one-particle representations which we need take the form~\eqref{eq:repr-bosonic} where now the representation coefficients are
\begin{equation}
\label{a_b_coefficients_after_the_limit}
a_m(\theta)= \sqrt{ h \Bigl|\sin \bigl(\frac{\pi m}{k} \bigr) \Bigr|} e^{\frac{\theta}{2}},\qquad
b_m(\theta)= \ e^{-i\frac{\pi m}{k}} g_m \sqrt{ h \Bigl| \sin \bigl(\frac{\pi m}{k} \bigr) \Bigr|} e^{-\frac{\theta}{2}} .
\end{equation}
where
\begin{equation}
g_m \equiv 
\begin{cases}
+1  \quad &\text{if} \hspace{9mm} 0<m<k \ \text{mod} \  2k \, ,\\
-1 \quad &\text{if} \quad -k<m<0  \ \text{mod} \  2k\, . 
\end{cases}
\end{equation}
As in~\cite{Frolov:2023lwd}, the solutions for $a_m$ and $b_m$ are chosen to realise the periodicity
\begin{equation}
\label{periodicity_on_the_coefficients_of_the_supercharges_massive_case}
a_{m \pm k}(\theta)=a_{m}(\theta)  \hspace{4mm},\hspace{4mm} b_{m \pm k}(\theta)=b_{m}(\theta).
\end{equation}
%\begin{subequations}
%\label{relations_on_coefficients_of_supercharges_after_the_limit}
%\begin{align}
%&|a_m(\theta)|^2 = \frac{\mu_m}{2} e^\theta,\\
%&|b_{m}(\theta)|^2 = \frac{\mu_m}{2} e^{-\theta},\\
%&a_m(\theta)b_m(\theta) = \frac{ih}{2} \bigl(e^{-2 \pi i\frac{m }{k}} -1 \bigr),\\
%&\bar{a}_m(\theta)\bar{b}_m(\theta) = -\frac{ih}{2} \bigl(e^{2 \pi i\frac{m }{k}} -1 \bigr).
%\end{align}
%\end{subequations}
The supercharges on double-particle states are determined by the following nontrivial coproduct
\begin{subequations}
\label{eq:limit_of_supercharge_Qm1m2}
\begin{align}
&\mathbf{q}_{m' , m''} (\theta', \theta'')= \mathbf{q}_{m'} (\theta') \otimes 1 + e^{-\frac{i \pi m'}{k}} \ \Sigma \otimes \mathbf{q}_{m''} (\theta'')\ ,
\\
\label{limit_of_supercharge_barQm1m2}
&\mathbf{s}_{m' , m''} (\theta', \theta'')= \mathbf{s}_{m'} (\theta') \otimes 1 + e^{\frac{i \pi m'}{k}} \ \Sigma \otimes \mathbf{s}_{m''} (\theta'') \ ,\\
&\mathbf{\tilde{q}}_{m' , m''} (\theta', \theta'')= \mathbf{\tilde{q}}_{m'} (\theta') \otimes 1 + e^{-\frac{i \pi m'}{k}} \ \Sigma \otimes \mathbf{\tilde{q}}_{m''} (\theta'')\ ,
\\
&\mathbf{\tilde{s}}_{m' , m''} (\theta', \theta'')= \mathbf{\tilde{s}}_{m'} (\theta') \otimes 1 + e^{\frac{i \pi m'}{k}} \ \Sigma \otimes \mathbf{\tilde{s}}_{m''} (\theta'') \ ,\\
&\mathbf{C}_{m' , m''} (\theta', \theta'')=\mathbf{C}_{m'} (\theta') \otimes 1 + e^{-\frac{2 i \pi m'}{k}}  \otimes \mathbf{C}_{m''} (\theta'')=\frac{i h}{2} \Bigl(e^{- \frac{2 i \pi}{k} (m'+m'')}-1 \Bigl) \ ,\\
&\mathbf{\bar{C}}_{m' , m''} (\theta', \theta'')=\mathbf{\bar{C}}_{m'} (\theta') \otimes 1 + e^{\frac{2 i \pi m'}{k}}  \otimes \mathbf{\bar{C}}_{m''} (\theta'')=-\frac{i h}{2} \Bigl(e^{\frac{2 i \pi}{k} (m'+m'')}-1 \Bigl) \ ,
\end{align}
\end{subequations}
where we denote by $\theta'$ and $\theta''$ (and $m'$ and $m''$) the rapidities (masses) of the first and second particle.
$\Sigma=(-1)^F$ corresponds to the fermion sign, in this case for the first particle. We notice that a shift $m' \to m' \pm k$ is equivalent to a change of sign of $\Sigma$, and therefore in the limit we can identify
\begin{equation}
\label{eq:k_periodicity_repres}
\rho^\B_m(\theta) \simeq \rho^\F_{m \pm k}(\theta) \,.
\end{equation}
This fact was early discussed in~\cite{Frolov:2023lwd}.
In particular, a right representation with mass $-m$ (and fermionic h.w.s.) is isomorphic to a left representation with mass $k-m$ (and bosonic h.w.s.). This identification is also valid at the level of the full model, up to a simple monodromy of the S~matrix~\cite{Frolov:2025uwz}.

%%%%%%%%%%%%
\subsection{Relativistic limit of the massive S-matrix}
\label{sec:general_structure_HWS_fusion}

The S~matrix describing the scattering of two massive particles was obtained in~\cite{Fontanella:2019ury,Frolov:2023lwd}.
If we focus on the scattering of particles in the representation $\rho^\B_{m'}(\theta')$ and $\rho^\B_{m''}(\theta'')$, then the Zamolodchikov--Faddeev (ZF) algebra is, introducing $\theta\equiv\theta'-\theta''$,
\begin{equation}
\label{ZF_algebra_LL_massive_sector_relativistic_limit}
\begin{split}
    &\phi^\B_{m'}(\theta') \phi^\B_{m''}(\theta'')=A_{m',m''}(\theta)\, \phi^\B_{m''}(\theta'') \phi^\B_{m'}(\theta') \, , \\
    &\phi^\B_{m'}(\theta') \varphi^\F_{m''}(\theta'')=B_{m',m''}(\theta)\, \varphi^\F_{m''}(\theta'') \phi^\B_{m'}(\theta') + C_{m',m''}(\theta)\, \phi^\B_{m''}(\theta'') \varphi^\F_{m'}(\theta') \, ,\\
    &\varphi^\F_{m'}(\theta') \phi^\B_{m''}(\theta'')=D_{m',m''}(\theta)\, \phi^\B_{m''}(\theta'') \varphi^\F_{m'}(\theta') + E_{m',m''}(\theta)\, \varphi^\F_{m''}(\theta'') \phi^\B_{m'}(\theta') \, ,\\
    &\varphi^\F_{m'}(\theta') \varphi^\F_{m''}(\theta'')=F_{m',m''}(\theta)\, \varphi^\F_{m''}(\theta'') \varphi^\F_{m'}(\theta') \, .
\end{split}
\end{equation}
In matrix form, this can be written as
\begin{equation}
\label{eq:ZFalgebraappendix}
S_{m' m''}(\theta)=
\begin{pmatrix}
A_{m', m''}(\theta) & 0 & 0 & 0\\
0 & C_{m', m''}(\theta) & D_{m', m''}(\theta) & 0\\
0 & B_{m', m''}(\theta) & E_{m', m''}(\theta) & 0\\
0 & 0 & 0 & F_{m', m''}(\theta)
\end{pmatrix} \ ,
\end{equation}
where the matrix coefficients, for any pair of masses $m', m'' \in (0, k)$, are
\begin{equation}
\label{eq:Smatrixelements-appendix}
\begin{aligned}
    &A_{m', m''}(\theta)= 1 \, ,\\
    &B_{m',m''}(\theta)= \frac{\sinh \Bigl(\frac{\theta}{2} - \frac{i \pi}{2 k} (m'-m'') \Bigr)}{\sinh \Bigl(\frac{\theta}{2} + \frac{i \pi}{2 k} (m'+m'') \Bigr)} \, ,\\
    &C_{m',m''}(\theta)= \frac{i \sqrt{\sin \bigl(m' \pi/k\bigr)} \sqrt{\sin \bigl(m'' \pi/k \bigr)}}{ \sinh \Bigl(\frac{\theta}{2} + \frac{i \pi}{2k} (m'+m'') \Bigr)} e^{\frac{i \pi}{2 k} (m'-m'')} \, ,\\
    &D_{m',m''}(\theta)= \frac{\sinh \Bigl(\frac{\theta}{2} + \frac{i \pi}{2 k} (m'-m'') \Bigr)}{\sinh \Bigl(\frac{\theta}{2} + \frac{i \pi}{2 k} (m'+m'') \Bigr)} \, ,\\
    &E_{m',m''}(\theta)=\frac{i \sqrt{\sin \bigl(m' \pi/k\bigr)} \sqrt{\sin \bigl(m'' \pi/k \bigr)}}{ \sinh \Bigl(\frac{\theta}{2} + \frac{i \pi}{2k} (m'+m'') \Bigr)} e^{-\frac{i \pi}{2 k} (m'-m'')} \, ,\\
    &F_{m',m''}(\theta)=- \frac{\sinh \Bigl(\frac{\theta}{2} - \frac{i \pi}{2 k} (m'+m'') \Bigr)}{\sinh \Bigl(\frac{\theta}{2} + \frac{i \pi}{2 k} (m'+m'') \Bigr)} \, .
\end{aligned}
\end{equation}
Note that we have conventionally set $A=1$.
The full S~matrix, whose elements we indicate by $\mathcal{A}$, $\mathcal{B}$, through $\mathcal{F}$, differs from the above by an  overall multiplicative scalar phase: the dressing factor. Introducing such a dressing factor we can write the full S-matrix as
\begin{equation}
\label{multiplication_between_dressing_phases_and_matrix_part_of_SLL}
\SM_{m' m''}(\theta)= \sigma_{m' m''} (\theta) S_{m' m''}(\theta).
\end{equation}
Of course, this factor needs to be added also to the RHS of each row in~\eqref{ZF_algebra_LL_massive_sector_relativistic_limit}.
As it happens for the S-matrix, we expect this dressing factor to be analytic in $\theta$ but not in $m'$ and $m''$.

The dressing factor must satisfy the following crossing equation
\begin{equation}
\label{first_crossing_equation_on_dressing_phases}
   \sigma_{m , m_3}(\theta) \sigma_{k-m , m_3}(\theta+i \pi) = \frac{\sinh \Bigl( \frac{\theta}{2}-\frac{i \pi}{2 k} (m-m_3) \Bigr)}{\sinh \Bigl( \frac{\theta}{2}-\frac{i \pi}{2 k} (m+m_3) \Bigr)} \, .
\end{equation}
A closed solution for this equation was originally found in~\cite{Fendley:1992dm} and expressed in~\cite{Frolov:2023lwd} in terms of the following building blocks
\begin{equation}
\label{R_function_definition}
     R(\theta)\equiv \frac{G(1- \frac{\theta}{2\pi i})}{G(1+ \frac{\theta}{2\pi i}) }\,,
\end{equation}
where $G(z)$ is the Barnes G-function.
The function~$R(\theta)$ obeys the properties
\begin{equation}
    R (-\theta)\,R (\theta)=1\,,\qquad
    [R(\theta^*)]^*\,R(\theta)=1\,,
\end{equation}
as well as the monodromy relations
\begin{equation}
\label{eq:R_function_monodromy}
    R(\theta-2\pi i) =i\,  \frac{\pi}{ \sinh{\tfrac{\theta}{2}}}\,R(\theta)\,,\qquad  R (\theta+\pi i) =   \frac{\cosh{\tfrac{\theta}{2}}}{\pi} \,R(\theta-\pi i)\,,
\end{equation}
The solution to the crossing equations can then be written in a closed form as follows
\begin{equation}
\label{eq:dressing_bosonicHWS}
  \sigma_{m' m''} (\theta)  \equiv \Phi_{m' m''}(\theta) \, \sigma^{\text{min}}_{m' m''} (\theta) \,.
\end{equation}
where
\begin{equation}
\label{minimal_sigma_ratio_of_R_functions}
\sigma^{\text{min}}_{m', m''}(\theta) =  \frac{R\left(\theta-\frac{i\pi(m'+m'')}{k}\right)\,R\left(\theta+\frac{i\pi(m'+m'')}{k}\right)}{R\left(\theta-\frac{i\pi(m'-m'')}{k}\right) \,R\left(\theta+\frac{i\pi(m'-m'')}{k}\right)} \,,
\end{equation}
and $\Phi_{m' m''}(\theta)$ is a ``CDD''~\cite{Castillejo:1955ed} factor satisfying the homogeneous crossing equation
\begin{equation}
\Phi_{m , m''}(\theta) \Phi_{k-m , m''}(\theta+i \pi)=1 \,.
\end{equation}
However, this CDD factor is important to reproduce the correct pole and fusion structure of the S~matrix; in the absence of poles, one could simply set $\Phi_{m' , m''}(\theta)=1$. 
We label the complete S-matrix elements (comprising the dressing factor) for the scattering of the highest and lowest weight states by
\begin{equation}
\begin{split}
&\mathcal{A}_{m', m''}(\theta)=\Phi_{m' m''}(\theta) \, \sigma^{\text{min}}_{m' m''} (\theta)\,,\\
&\mathcal{F}_{m', m''}(\theta)= - \frac{\sinh \Bigl(\frac{\theta}{2} - \frac{i \pi}{2 k} (m'+m'') \Bigr)}{\sinh \Bigl(\frac{\theta}{2} + \frac{i \pi}{2 k} (m'+m'') \Bigr)} \Phi_{m' m''}(\theta) \, \sigma^{\text{min}}_{m' m''} (\theta) \,.
\end{split}
\end{equation}

In the remainder of this appendix, we first present the solution proposed in~\cite{Fendley:1992dm} by Fendley and Intriligator. It is closely related to the $\AdSST$ solution, as we will see below, and in fact it gives almost on the nose the solution for $\AdSSSS$ too, as discussed in the main body of this paper.

%%%%%%%%%%%%
\subsection{Fendley-Intriligator solution}
\label{app:FI_solution}

In the model studied in~\cite{Fendley:1992dm}, $k$ is an integer and the masses $m'$ and $m''$ take values in $\{1, \dots, k-1 \}$. The additional CDD factor takes the following form
\begin{equation}
\label{eq:CDD_proposal_old}
\Phi_{m' m''}(\theta)= \Bigl[\frac{m'+m''}{k} \Bigl]_\theta \, \Bigl[ \frac{m'+m'' -2}{k}\Bigl]^2_\theta \, \dots \Bigl[\frac{|m' - m''|+2}{k}\Bigl]_\theta^2 \, \Bigl[\frac{|m'-m''|}{k}\Bigl]_\theta \,,
\end{equation}
where
\begin{equation}
\label{eq:[x]_bb}
\Bigl[\frac{m}{k} \Bigl]_{\theta} \equiv \frac{\sinh \left( \frac{\theta}{2} + \frac{i \pi m}{2k} \right)}{\sinh \left( \frac{\theta}{2} - \frac{i \pi m}{2k} \right)} \,.
\end{equation}
This solution leads to a bound-state pole for the scattering of h.w.s.\ when the masses obey $0<m'+m''<k$:  more precisely, there is a simple pole in $\mathcal{A}_{m' m''}$  at $\theta=\frac{i \pi}{k} (m'+m'')$. For $k<m'+m''<2k$ this pole goes outside the physical strip $(0 , i \pi)$ and a pole $\theta=\frac{i \pi}{k} (2k-m'-m'')$ enters in the physical strip for the element $\mathcal{F}_{m' m''}$. This corresponds to the scattering of two l.w.s.. It is possible to show that the bootstrap closes with exactly $k-1$ massive representations.
The solution by Fendly and Intriligator can then be expressed in the following compact form
\begin{equation}
\label{eq:app_FI_S_mat}
\mathcal{S}^{\text{FI}}_{m', m''}(\theta)= \Phi_{m' m''}(\theta) \,\sigma^{\text{min}}_{m' m''}(\theta) \, S_{m', m''}(\theta) \,,
\end{equation}
where the minimal dressing factor is the one in~\eqref{minimal_sigma_ratio_of_R_functions}.

%%%%%%%%%%%%
\subsection{Solution for \texorpdfstring{$\AdSST$}{AdS3xS3xT4}}

The S~matrix proposed in~\cite{Frolov:2023lwd} for the relativistic limit of the background $\AdSST$ acts on four dimensional representations of the form $\rho_{m}^\B(\theta) \otimes \rho_{m}^\B(\theta)$.
This is a substantial difference compared with the one in~\cite{Fendley:1992dm}, which acts instead on representations $\rho_{m}^\B(\theta)$. The matrix structure and crossing equations of the former are therefore the square of those discussed by Fendley and Intriligator. However, the pole structure does not ``square'', but rather the CDD factor has the same form as the one of Fendley and Intriligator. All in all, the S~matrix proposed in~\cite{Frolov:2023lwd} is
\begin{equation}
\mathcal{S}^{\AdSST}_{m', m''}(\theta)= \Phi_{m' m''}(\theta) \,(\sigma^{\text{min}}_{m' m''}(\theta))^2 \, \left(S_{m', m''}(\theta) \otimes S_{m', m''}(\theta) \right) \,.
\end{equation}

%%%%%%%%%%%%

\section{Example: the case \texorpdfstring{$k_1=3$, $k_2=6$}{k1=3, k2=6}}
\label{app:example}

In this appendix, we provide an example of the relativistic limit of the worldsheet S~matrix of mixed-flux $\AdSSSS$ when $k_1=3$ and $k_2=6$. Recalling the condition~\eqref{eq:alphakmap}, this means 
\begin{equation}
k=\frac{k_1k_2}{k_1+k_2}=2\,,\qquad
\alpha=\frac{k_2}{k_1+k_2}= \frac{2}{3} \,, \qquad 1-\alpha=\frac{k_1}{k_1+k_2}= \frac{1}{3} \,.
\end{equation}
We describe both the S~matrix arising from assuming $\alpha$ and $(1-\alpha)$ to be the minimal masses of the theory (this S~matrix corresponds to the relativistic limit of the solution proposed in~\cite{Frolov:2025syj}) and the case in which they are bound states made of a fictitious minimal lighter particle. We split this second case into options A and B, according to section~\ref{sec:fictitious_min_part}.

%%%%%%%%%%%%%%%%%%%%%%%%%%
\subsection{\texorpdfstring{$\alpha$}{alpha} and \texorpdfstring{$1-\alpha$}{1 - alpha} as fundamental particles}
\label{sec:app_separate_sectors}

Let us first work with the assumption that $\alpha$ and $1-\alpha$ are separate sectors, which means that fusing two particles of mass $1-\alpha$ does not generate a particle of type $\alpha$, even though the bound state has the same mass as $\alpha$.

%%%%
\paragraph{The $(\alpha,\,\alpha)$ sector.}
For the scattering of particles of type $\alpha$ we have
\begin{equation}
\begin{split}
&\mathcal{A}^{\alpha \alpha}_{Q' Q''}(\theta)=\Phi^{\alpha \alpha}_{Q' Q''}(\theta) \, \sigma^{\alpha \alpha, \text{min}}_{Q' Q''} (\theta)\,,\\
&\mathcal{F}^{\alpha \alpha}_{Q' Q''}(\theta)= - \frac{\sinh \Bigl(\frac{\theta}{2} - \frac{i \pi}{6} (Q'+Q'') \Bigr)}{\sinh \Bigl(\frac{\theta}{2} + \frac{i \pi}{6} (Q'+Q'') \Bigr)} \Phi^{\alpha \alpha}_{Q' Q''}(\theta) \, \sigma^{\alpha \alpha, \text{min}}_{Q' Q''} (\theta) \,,\\
&Q', Q''= 1, \, 2 \,.
\end{split}
\end{equation}
In this case, the CDD factor is given by
\begin{equation}
\Phi^{\alpha \alpha}_{Q' Q''}(\theta)= \Bigl[ \frac{Q'+Q''}{3} \Bigl]_\theta \, \Bigl[ \frac{Q'+Q'' -2}{3} \Bigl]^2_\theta \, \dots \Bigl[\frac{|Q' - Q''|+2}{3} \Bigl]_\theta^2 \, \Bigl[\frac{|Q' - Q''|}{3}\Bigl]_\theta \,.
\end{equation}
This leads to
\begin{equation}
\Phi^{\alpha \alpha}_{11}(\theta)=\Phi^{\alpha \alpha}_{22}(\theta)=\Bigl[ \frac{2}{3} \Bigl]_\theta\,, \qquad \Phi^{\alpha \alpha}_{21}(\theta)=\Phi^{\alpha \alpha}_{12}(\theta)=- \Bigl[\frac{1}{3} \Bigl]_\theta \,.
\end{equation}

%%%%
\paragraph{The $(1-\alpha, \,1-\alpha)$ sector.}

In this sector, we have
\begin{equation}
\begin{split}
&\mathcal{A}^{1-\alpha, 1-\alpha}_{Q' Q''}(\theta)=\Phi^{1-\alpha, 1-\alpha}_{Q' Q''}(\theta) \, \sigma^{1-\alpha, 1-\alpha, \text{min}}_{Q' Q''} (\theta)\,,\\
&\mathcal{F}^{1-\alpha, 1-\alpha}_{Q' Q''}(\theta)= - \frac{\sinh \Bigl(\frac{\theta}{2} - \frac{i \pi}{12} (Q'+Q'') \Bigr)}{\sinh \Bigl(\frac{\theta}{2} + \frac{i \pi}{12} (Q' + Q'') \Bigr)} \Phi^{1-\alpha, 1-\alpha}_{Q' Q''}(\theta) \, \sigma^{1-\alpha, 1-\alpha, \text{min}}_{Q' Q''} (\theta) \,,\\
&Q', Q''= 1,\, 2,\,3,\, 4,\,5\,.
\end{split}
\end{equation}
The CDD factor is given by
\begin{equation}
\Phi^{1-\alpha, 1-\alpha}_{Q' Q''}(\theta)= \Bigl[ \frac{Q'+Q''}{6} \Bigl]_\theta \, \Bigl[\frac{Q'+Q'' -2}{6} \Bigl]^2_\theta \, \dots \Bigl[\frac{|Q' - Q''|+2}{6} \Bigl]_\theta^2 \, \Bigl[\frac{|Q'-Q''|}{6} \Bigl]_\theta \,.
\end{equation}
Using that
\begin{equation}
\Bigl[\frac{6+Q}{6} \Bigl]_\theta \Bigl[\frac{6-Q}{6} \Bigl]_\theta=1 \,,
\end{equation} 
then the CDDs can be written explicitly as
\begin{equation}
\begin{split}
&\Phi^{1-\alpha, 1-\alpha}_{1 1}(\theta)=\Phi^{1-\alpha, 1-\alpha}_{5 5}(\theta) = [2/6]_\theta \,, \qquad \qquad  \ \ \Phi^{1-\alpha, 1-\alpha}_{2 1}(\theta)= \Phi^{1-\alpha, 1-\alpha}_{5 4}(\theta)=  [3/6]_\theta [1/6]_\theta \,,\\
&\Phi^{1-\alpha, 1-\alpha}_{3 1}(\theta)=\Phi^{1-\alpha, 1-\alpha}_{5 3}(\theta)= [4/6]_\theta [2/6]_\theta\,, \qquad  \, \Phi^{1-\alpha, 1-\alpha}_{4 1}(\theta)= \Phi^{1-\alpha, 1-\alpha}_{5 2}(\theta)=[5/6]_\theta [3/6]_\theta \,,\\
&\Phi^{1-\alpha, 1-\alpha}_{2 2}(\theta)= \Phi^{1-\alpha, 1-\alpha}_{4 4}(\theta)=[4/6]_\theta [2/6]^2_\theta\,, \qquad \ \Phi^{1-\alpha, 1-\alpha}_{3 3}(\theta)= -[4]^2_\theta [2]^2_\theta \,,\\
&\Phi^{1-\alpha, 1-\alpha}_{4 2}(\theta)= [6/6]_\theta [4/6]^2_\theta [2/6]_\theta \,, \hspace{25mm} \Phi^{1-\alpha, 1-\alpha}_{5 1}(\theta)= -[4/6]_\theta \,,\\
&\Phi^{1-\alpha, 1-\alpha}_{3 2}(\theta)=\Phi^{1-\alpha, 1-\alpha}_{4 3}(\theta)= [5/6]_\theta [3/6]^2_\theta [1/6]_\theta \,,
\end{split}
\end{equation}
with the remaining elements obtained by parity.

%%%%
\paragraph{The $(\alpha, \,1-\alpha)$ sector.}

As discussed in section~\ref{sec:bound_st_Smat}, we assume no poles in the scattering of particles of different masses, $\alpha$ and $1-\alpha$. The mixed-mass relativistic S-matrix then takes the following form: 
\begin{equation}
\begin{split}
&\mathcal{A}^{\alpha, 1-\alpha}_{11}(\theta)= \sigma^{\alpha, 1-\alpha, \text{min}}_{1 1} (\theta)\,,\\
&\mathcal{F}^{\alpha, 1-\alpha}_{11}(\theta)= - \frac{\sinh \Bigl(\frac{\theta}{2} - \frac{i \pi}{2} (\frac{1}{3}+\frac{1}{6}) \Bigr)}{\sinh \Bigl(\frac{\theta}{2} + \frac{i \pi}{2} (\frac{1}{3}+\frac{1}{6}) \Bigr)}  \, \sigma^{\alpha, 1-\alpha, \text{min}}_{1 1} (\theta) \,.
\end{split}
\end{equation}
Notice that we set the CDD factors to one, since we do not require the presence of poles.
Applying fusion, we generate 
\begin{equation}
\begin{split}
&\mathcal{A}^{\alpha, 1-\alpha}_{Q' Q''}(\theta)= \sigma^{\alpha, 1-\alpha, \text{min}}_{Q' Q''} (\theta)\,,\\
&\mathcal{F}^{\alpha, 1-\alpha}_{Q', Q''}(\theta)= - \frac{\sinh \Bigl(\frac{\theta}{2} - \frac{i \pi}{2} (\frac{Q'}{3}+\frac{Q''}{6}) \Bigr)}{\sinh \Bigl(\frac{\theta}{2} + \frac{i \pi}{2} (\frac{Q'}{3}+ \frac{Q''}{6}) \Bigr)}  \, \sigma^{\alpha, 1-\alpha, \text{min}}_{Q' Q''} (\theta) \,,\\
&Q'=1,\,2\,, \qquad  Q''= 1,\, 2,\,3,\, 4,\,5\,.
\end{split}
\end{equation}
Indeed, the minimal dressing factor fuses straightforwardly. For $\frac{Q'}{3}+\frac{Q''}{6}>1$ the term
$$
\frac{\sinh \Bigl(\frac{\theta}{2} - \frac{i \pi}{2} (\frac{Q'}{3}+ \frac{Q''}{6}) \Bigr)}{\sinh \Bigl(\frac{\theta}{2} + \frac{i \pi}{2} (\frac{Q'}{3}+\frac{Q''}{6}) \Bigr)}
$$
features a pole at $2 i \pi - i \pi (Q'/3 + Q''/6)$. However, this pole is compensated by a zero in the minimal dressing factor,%
\footnote{\label{foot:R}
Note that the function $R(z)$ has poles located at $
z=-2 \pi i n$ and zeros at $z=+2 \pi i n$ for $n=1, 2, 3, \dots\,$.}
and both $\mathcal{A}$ and $\mathcal{F}$ have no poles in the physical strip. Because of this, we do not need to require the existence of any additional propagating bound states. The spectrum closes with the following representations. From the $\alpha$ sector we have
\begin{equation}
\alpha: \qquad \rho_{2/3}^\B(\theta)\,, \ \rho_{4/3}^\B(\theta) \,.
\end{equation}
From the $1-\alpha$ sector, we have
\begin{equation}
1-\alpha: \qquad \rho_{1/3}^\B(\theta)\,, \ \rho_{2/3}^\B(\theta) \,, \ \rho_1^\B(\theta)\,, \ \rho_{4/3}^\B(\theta)\,, \ \rho_{5/3}^\B(\theta) \,.
\end{equation}
The two sectors share some representations with the same mass. However, the S-matrix elements of these representations are different, keeping track of the fact that the representations originate from two different three-spheres, distinguishable in principle using the $\mathfrak{u}(1)$ charges in~\eqref{eq:u1charges}. Therefore, they genuinely correspond to different particles.

%%%%%%%%%%%%%%%%%%%%%%%%%%
\subsection{A common minimal fundamental particle: option B}
\label{app:optB}

We start considering the case in which the fictitious particle of minimal mass is obtained by taking the common denominator between $1/k_1$ and $1/k_2$, which is option~B in~\eqref{eq:kzero}.%
\footnote{Hopefully, the reader will forgive the reverse alphabetical order.}
The mcm between $k_1=3$ and $k_2=6$ is 
\begin{equation}
k_0=6 \,.
\end{equation}
We observe that the particle of mass $m=\alpha$ is a bound state made of $k_0/k_1=2$ particles of mass $(1-\alpha)$, which is assumed to be our lightest fundamental particle of minimal mass.
We can write a closed formula for the S-matrix elements associated with the scattering of h.w.s. and l.w.s. as follows
\begin{equation}
\begin{split}
&\mathcal{A}_{Q', Q''}(\theta)=\Phi_{Q' Q''}(\theta) \, \sigma^{\text{min}}_{Q' Q''} (\theta)\,,\\
&\mathcal{F}_{Q', Q''}(\theta)= - \frac{\sinh \Bigl(\frac{\theta}{2} - \frac{i \pi}{2 k_0} (Q'+Q'') \Bigr)}{\sinh \Bigl(\frac{\theta}{2} + \frac{i \pi}{2 k_0} (Q'+Q'') \Bigr)} \Phi_{Q' Q''}(\theta) \, \sigma^{\text{min}}_{Q' Q''} (\theta) \,.
\end{split}
\end{equation}
where
\begin{equation}
\sigma^{\text{min}}_{Q', Q''}(\theta) =  \frac{R\left(\theta-\frac{i\pi(Q'+Q'')}{k_0}\right)\,R\left(\theta+\frac{i\pi(Q'+Q'')}{k_0}\right)}{R\left(\theta-\frac{i\pi(Q'-Q'')}{k_0}\right) \,R\left(\theta+\frac{i\pi(Q'-Q'')}{k_0}\right)} \,,
\end{equation}
\begin{equation}
\Phi_{Q' Q''}(\theta)= \Bigl[\frac{Q'+Q''}{k_0} \Bigl]_\theta \, \Bigl[\frac{Q'+Q'' -2}{k_0}\Bigl]_\theta^2 \, \dots \Bigl[\frac{|Q' - Q''|+2}{k_0}\Bigl]_\theta^2 \, \Bigl[\frac{|Q'-Q''|}{k_0}\Bigl]_\theta \,,
\end{equation}
and $Q', \, Q''=1, \, \dots, \, 5$.
Let us note that the minimal dressing factor $\sigma^{\text{min}}_{Q', Q''}(\theta)$ has no poles in the physical strip $(0, i \pi)$ for all integer values $Q', Q'' \in \{1, \dots, 5\}$.

\paragraph{The case $Q'+Q''<k_0$.}

If $Q'+Q''<k_0$ then the CDD factor $\Phi_{Q' Q''}(\theta)$ has a pole at
\begin{equation}
\theta=\frac{i \pi}{k_0} (Q' + Q'')\,,
\end{equation}
due to the building block $[\frac{Q'+Q''}{k_0} ]_\theta$. This is a pole in the element $\mathcal{A}_{Q', Q''}(\theta)$. The element $\mathcal{F}_{Q', Q''}(\theta)$ is instead regular due to a zero in~$F_{Q', Q''}(\theta)$. We conclude that for $Q'+Q''<k_0$, the bound states are obtained by fusing the highest weight states as follows
\begin{equation}
\phi_{Q''}^\B\left(\theta - \frac{i \pi}{k_0} Q'\right)\ \phi_{Q'}^\B\left(\theta + \frac{i \pi}{k_0} Q'' \right) = \phi_{Q'+Q''}^\B\left(\theta\right) \,, 
\end{equation}
where $\phi_{Q}\left(\theta\right)$ indicate creation operators of the Zamolodchikov--Faddeev algebra.

\paragraph{The case $Q'+Q''>k_0$.}
The minimal dressing factor~$\sigma^{\text{min}}_{Q'Q''}$ has a simple zero at
\begin{equation}
\label{eq:zeron1plusn2big6}
\theta=\frac{i \pi}{k_0} (2k_0-Q'-Q'') \,,
\end{equation}
see footnote~\ref{foot:R}, and for $Q'+Q''>k_0$ this zero is inside the physical strip.
However, $\mathcal{A}_{Q' Q''}(\theta)=\Phi_{Q' Q''}(\theta)\sigma^{\text{min}}_{Q'Q''}$ is regular due to a simple pole in the CDD factor~$\Phi_{Q' Q''}(\theta)$ at the same location.%
\footnote{To work out the pole structure of the CDD factor it is useful to note the identity
$[{x}/{k_0}]_\theta \cdot [{(2 k_0-x)}/{k_0} ]_\theta=1 $ which allows to telescope many of the terms in the product. One finds  that, for $k_0<Q'+Q''<2k_0$ we have $\Phi_{Q',Q''} (\theta)=[{(2k_0-Q' - Q'')}/{k_0}]_{\theta}$ $[{(2 k_0- Q' - Q'' -2)}/{k_0} ]^2_{\theta}$ $\cdots$ $[{(|Q'-Q''|+2)}/{k_0} ]^2_{\theta}$ $[{|Q'-Q''|}/{k_0} ]_{\theta}$.
}
Conversely, $\mathcal{F}_{Q' Q''}(\theta)$ has a simple pole at~\eqref{eq:zeron1plusn2big6} coming from $F_{Q' Q''}(\theta)$. Because of this, for $Q'+Q''>k_0$ fusion is realised between lower weight states and we should identify
\begin{equation}
\varphi_{Q''}^\F\left(\theta - \frac{i \pi(k_0-Q')}{k_0} \right)\ \varphi_{Q'}^\F\left(\theta + \frac{i \pi(k_0-Q'')}{k_0}  \right) = \varphi_{Q'+Q''}^\B(\theta) = \varphi_{Q'+Q''-k_0}^\F(\theta) \,,
\end{equation}
in the ZF algebra.
In this way, fusion closes on representations with $(\phi^\B_Q,\varphi^\F_Q)$ with $Q=1,\dots, k_0-1$, without ever generating new representations with $Q>k_0$.

In conclusion, the spectrum closes onto the following representations
\begin{equation}
\rho_{1/3}^\B(\theta)\,, \ \rho_{2/3}^\B(\theta) \,, \ \rho_1^\B(\theta)\,, \ \rho_{4/3}^\B(\theta)\,, \ \rho_{5/3}^\B(\theta) \,,
\end{equation}
and there is exactly one representation for each different mass. The representation of minimal mass is the one with $m=1-\alpha=1/3$. In this picture, the two three-spheres share common representations and a particle of type $\alpha$ is identified with a bound state made of two particles of type $1-\alpha$.

%%%%%%%%%%%%%%%%%%%%%%%%%%
\subsection{A common minimal fundamental particle: option A}
\label{app:optA}

If we work with option A, then we require
\begin{equation}
k_0=k_1 \cdot k_2 = 18 \,.
\end{equation}
The bound state numbers now run over the values
\begin{equation}
Q', Q''=1, \, 2, \, \dots \, ,17 \,.
\end{equation}
There is then a large number of representations and poles introduced by the CDD factor $\Phi_{Q' Q''}$. Particles of masses $1-\alpha$ and $\alpha$ are bound states obtained by fusing  $k_1$ and $k_2$ times the particle of minimal mass:
\begin{equation}
    \begin{split}
    &2h \sin \frac{\pi}{k_0} \xrightarrow{\text{fusing} \ k_1-\text{times}}
       2h \sin \frac{k_1}{k_0} \pi=2h \sin \frac{\pi}{k_2}= 2h \sin \frac{(1-\alpha) \pi}{k} \,,\\
    &2h \sin \frac{\pi}{k_0} \xrightarrow{\text{fusing} \ k_2-\text{times}}
       2h \sin \frac{k_2}{k_0} \pi=2h \sin \frac{\pi}{k_1}= 2h \sin \frac{\alpha \pi}{k} \,.
    \end{split}
\end{equation}
The CDD factor for the scattering of $\alpha$ particles is given by
\begin{equation}
\Phi_{k_2 k_2}(\theta) = \Bigl[\frac{2 k_2}{k_0} \Bigl]_\theta \, \Bigl[\frac{2 k_2-2}{k_0}\Bigl]_\theta^2 \dots \Bigl[\frac{2}{k_0}\Bigl]_\theta^2 \, \Bigl[\frac{0}{k_0}\Bigl] = \Bigl[\frac{12}{18}\Bigl]_\theta \, \Bigl[\frac{10}{18}\Bigl]_\theta^2 \dots \Bigl[\frac{2}{18}\Bigl]_\theta^2 \, [0] \,.
\end{equation}
This includes a large number of poles that must be associated with fictitious particles propagating on-shell in Feynman diagrams of Coleman-Thun type~\cite{Coleman:1978kk}. Since these poles are present in the scattering of physical particles, we cannot quite say that the particle of minimal mass we started with is fictitious. Indeed, by fusion it generates a large number of `fictitious' bound states that affect in a nontrivial way the S-matrices of $\alpha$ and $1-\alpha$ particles. These bound states can then be detected in the physical scattering.

In general, the same argument is valid with the option~B, even though in this case, the number of `fictitious' bound states is reduced. If these poles exist in the relativistic limit, then they should also be present before the limit. This seems to be inconsistent with perturbative computations of the S~matrix in the near-BMN expansion, where there is no sign of these additional poles.
This leads to the belief that the starting assumption of having a fictitious particle of minimal mass should be incorrect, and is instead preferable to consider $\alpha$ and $1-\alpha$ to be our particles from which to generate all the bound states.

%%%%%%%%%%%%%%%%%%%%%%
\bibliographystyle{JHEP}
\bibliography{refs}

\end{document}